\documentclass[twocolumn, apj, numberedappendix, tighten, twocolappendix]{openjournal}

\usepackage{graphicx}
\usepackage{dcolumn}
\usepackage{bm}
\usepackage{cprotect}
\usepackage{rviewport}

\def\apjl{{ApJ}}                
\def\apjs{{ApJS}}               
\def\ao{{Appl.~Opt.}}           
\def\aap{{A\&A}}                

\def\mnras{Mon.~Not.~Roy.~Astron.~Soc.}             
\def\prd{{Phys.~Rev.~D}}        
\def\physrep{{Phys.~Rep.}}   

\newcommand{\beq}{\begin{equation}}
\newcommand{\beqa}{\begin{eqnarray}}
\newcommand{\eeq}{\end{equation}}
\newcommand{\eeqa}{\end{eqnarray}}

\newcommand{\simgt}{\lower.5ex\hbox{$\; \buildrel > \over \sim \;$}}
\newcommand{\simlt}{\lower.5ex\hbox{$\; \buildrel < \over \sim \;$}}

\newcommand{\bd}[1]{\mbox{\boldmath $#1$}}

\usepackage[usenames,dvipsnames]{xcolor}

\newcommand{\ms}[1]{{\color{black} #1}}

\usepackage{amsmath,amssymb,natbib,latexsym,times}
\usepackage[hidelinks]{hyperref}

\begin{document}


\title{Neural style transfer of weak lensing mass maps}

\author{Masato Shirasaki}
  \email{masato.shirasaki@nao.ac.jp}
\affiliation{%
National Astronomical Observatory of Japan (NAOJ), National Institutes of Natural Science, Mitaka, Tokyo 181-8588, Japan
}%
\affiliation{
The Institute of Statistical Mathematics,
Tachikawa, Tokyo 190-8562, Japan
}

\author{Shiro Ikeda}
\affiliation{
The Institute of Statistical Mathematics,
Tachikawa, Tokyo 190-8562, Japan
}
\affiliation{
Department of Statistical Science, School of Multidisciplinary Sciences, Graduate University for Advanced Studies, 10-3 Midori-cho,
Tachikawa, Tokyo 190-8562, Japan
}
\affiliation{
Kavli Institute for the Physics and Mathematics of the Universe, The University of Tokyo, 5-1-5 Kashiwanoha, Kashiwa, 277-8583,
Japan
}

\date{\today}

\begin{abstract}
We propose a new generative model of projected cosmic mass density maps inferred from weak gravitational lensing observations of distant galaxies (weak lensing mass maps).
We construct the model based on a neural style transfer so that it can transform Gaussian weak lensing mass maps into deeply non-Gaussian counterparts 
as predicted in ray-tracing lensing simulations.
We develop an unpaired image-to-image translation method with Cycle-Consistent Generative Adversarial Networks (Cycle GAN), which learn efficient mapping 
from an input domain to a target domain.
Our model is designed to enjoy important advantages;
it is trainable with no need for paired simulation data, 
flexible to make the input domain visually meaningful,
and expandable to rapidly-produce a map with a larger sky coverage than training data without additional learning.
Using 10,000 lensing simulations, \ms{we find that appropriate labeling of training data based on field variance allows the model to reproduce a correct scatter in summary statistics for weak lensing mass maps.}
Compared with a popular log-normal model, our model improves in predicting the statistical natures of three-point correlations and local properties of rare high-density regions.
We also demonstrate that our model enables us to produce a continuous map 
with a sky coverage of $\sim166\, \mathrm{deg}^2$ but 
similar non-Gaussian features to training data covering $\sim12\, \mathrm{deg}^2$
in a GPU minute.
Hence, our model can be beneficial to massive productions of synthetic weak lensing mass maps, which is of great importance in future precise real-world analyses.
\end{abstract}

\maketitle

\section{Introduction}

General relativity predicts that an apparent image of a single distant astrophysical source should be distorted by gravitational lensing effects. 
The distortion is governed by the gravitational potential of intervening mass (lens)
and the geometry between the lens and source objects \citep{Bartelmann:1999yn}.
Extracting coherent distortions over a large number of sources allows us 
to infer large-scale gravitating mass distributions of the Universe 
in an unbiased manner.
The lensing effect caused by the large-scale structure is commonly referred to as the weak lensing effect, whereas the mass distributions inferred from the weak lensing effect are called weak lensing mass maps in the literature.
Statistical analyses of weak lensing mass maps can provide vital information of gravitational clumping of dark matter and cosmic expansion (see, \citet{Kilbinger:2014cea} and \citet{Mandelbaum:2017jpr} for reviews).

The two-point angular correlation function of shapes of 
sources, or its Fourier-space counterpart known as the power spectrum, 
are commonly used to characterize weak lensing mass maps. 
Although the power spectrum provides a complete statistical description 
of a random Gaussian field, numerical simulations of weak lensing effects 
have shown that weak lensing mass maps have non-Gaussian properties.
Hence the power spectrum alone cannot fully describe a weak lensing mass map 
\citep{Jain:1999ir,White:1999xa}. 
Various statistical methods have been proposed to study the non-Gaussian features of weak lensing mass maps (see \citet{Euclid:2023uha} and references therein for a list of the statistics proposed so far).
At present, there are no unique methods to extract all the information imprinted in weak lensing mass maps. 

A generative model of weak lensing mass maps plays an essential role 
in modern weak lensing analyses. 
It is highly demanded in practice for the reasons below;

\begin{itemize}
    \item Valid analytic approaches are yet unknown to derive cosmological dependence of most of the non-Gaussian statistics proposed so far. An ensemble average of a given non-Gaussian statistic over a set of generated weak lensing mass maps is commonly used to study its cosmological dependence. 
    \item A precise estimation of statistical errors in a given lensing statistic is crucial in cosmological inference \citep{Dodelson:2013uaa}. 
    A strong non-Gaussianity in the weak lensing mass maps and several observational effects make it difficult to estimate the statistical errors in an analytic way. The best practice is to run a set of numerical simulations with the relevant physics to lensing analyses and to create realistic synthetic weak lensing data by including various observational effects with appropriate recipes \citep{Shirasaki:2019gya}.    
    \item Blind analyses of weak lensing mass maps become a standard procedure to remove any human-based systematic errors as some cosmological tensions have been claimed (see \citet{Abdalla:2022yfr} for a review of the cosmological tensions). 
    The current blinding scheme of weak lensing data is based on 
    changes of observed galaxy shapes with a single number, introducing modulations in the amplitude of lensing power spectrum \citep{KiDS:2020suj,DES:2021vln,Dalal:2023olq,Li:2023tui}. More sophisticated blinding is required to test the accuracy of a theoretical template for a given lensing statistic. A blinded challenge, analyzing generated weak lensing mass maps by making input cosmological parameters unknown, is among the most effective approaches for this purpose (see \citet{Nishimichi:2020tvu} for an example in galaxy clustering data). A fast precise generative model of weak lensing mass maps enables one to perform the blinded challenge in a user-friendly manner.
    \item Field-level cosmological inference of weak lensing mass maps receives plenty of attention from the community \citep{Porqueres:2021clw, Boruah:2022lsu, Dai:2022dso} because it can have no loss of cosmological information in principle. Reconstructions of weak lensing mass maps within a Bayesian framework are also an important subject in modern astronomy even for a fixed cosmological model, whereas the reconstruction from noisy galaxy imaging data often requires a training dataset of noiseless weak lensing mass maps \citep{Shirasaki:2018thk, Jeffrey:2019fag, Fiedorowicz:2021clg, Remy:2022ixn}. A rapid but realistic generator of weak lensing mass maps is fundamental to the field-level inference and Bayesian reconstruction as a matter of course. 
\end{itemize}

Ray-tracing simulations of weak lensing effects are among the most robust generative models thus far, but those need multiple outputs 
from cosmological simulations of large-scale structures.
Because of its computational cost, the ray-tracing simulation can typically provide 100-10,000 realizations of weak lensing mass maps on a yearly basis \citep{Sato:2009ct,Dietrich:2009jq,2012MNRAS.426.1262H,Liu:2017now,Takahashi:2017hjr,Harnois-Deraps:2018bcv,Harnois-Deraps:2019rsd,Shirasaki:2019wxk,Osato:2020sxo,Kacprzak:2022pww,Hadzhiyska:2023fic,Ferlito:2023gum}.
To reduce the computational cost, previous studies have proposed that the time-consuming cosmological N-body simulation can be replaced with some alternatives.
The proposals include the use of multivariate log-normal models \citep{Taruya:2002vy, Xavier:2016elr, Makiya:2020iai} or its advanced versions \citep{Yu:2016qoq, Tessore:2023zyk}, an approximate gravity solver \citep{Izard:2017kma, Sgier:2018soj, Bohm:2020ilt}, and halo-based models \citep{Giocoli:2017nnq, Giocoli:2020pfb}.
Ray-tracing simulations with the alternative methods drastically shorten computational time with a fixed computing environment, while they have limited predictive power for non-Gaussian observables in general.

Image processing with neural networks provides 
another interesting approach for the generative model of weak lensing effects.
Motivated by recent successes in deep learning, 
there exist several generative models of weak lensing mass maps based on 
deep-learning neural networks, including Generative Adversarial Networks (GANs) 
\citep{2019ComAC...6....1M, Perraudin:2020gig, Tamosiunas:2020rvw},
Score-based Diffusion Models \citep{Remy:2022ixn},
and Normalizing Flow \citep{Dai:2022dso}.
Those deep-learning-based generators commonly produce a new weak lensing mass map from a given set of random numbers, and the generators learn the relation between the mass map and latent random numbers from a large set of training data.
Because a field of view and the number of pixels in the training data has to be set a priori, the generators are not able to produce weak lensing mass maps 
at different sky coverage and angular resolution from the training setup.
This is a weak point of the deep-learning-based generative models
as ongoing and future galaxy imaging surveys aim at measuring weak lensing effects over a sky coverage of $1000\, \mathrm{deg}^2$ or larger.
In general, it becomes difficult to train a given neural 
network as the size of output data increases, 
because the number of trainable parameters in the network increases dramatically. 

In this paper, we propose a new generative model of weak lensing mass maps based on a deep-learning method, which is designed to be able to increase a field of view without additional training.
We train the neural network translating a latent image 
into a target image of a non-Gaussian weak lensing mass map.
We here set the latent image to a multivariate Gaussian map 
with the power spectrum being the same as the target counterpart.
The network will be trained to learn the translation between the Gaussian and non-Gaussian mass maps in a given field of view within the framework of 
Cycle-Consistent Generative Adversarial Networks (Cycle GAN) \citep{CycleGAN2017}.
After training, we can extend a sky coverage of fake mass maps 
by (i) producing a large-sky Gaussian map from the given power spectrum,
(ii) dividing the Gaussian map into subregions,
(iii) performing the learned translation on a patch-by-patch basis,
and (iv) combining the non-Gaussian maps on all the subregions.
This procedure allows our generator to produce a lensing mass map with an arbitrary sky coverage while not degrading the angular resolution compared to the training datasets.
Note that our training data does not require a large sky coverage, 
enabling the neural network to learn small-scale non-Gaussian features in a weak lensing mass map in a cost-efficient manner. 
Also, our approach enables us to rapidly increase the number of realistic lensing simulations with low costs of producing a new set of the Gaussian mass maps as long as we thoroughly train the Cycle GAN.

Similar approaches based on an image-to-image translation have been explored in \citet{Han:2021unz}
and \citet{Piras:2022dgt}.
A distinct difference from the previous studies is that our training does not need a paired or aligned set of images for the image-to-image translation.
This is achieved by recent progress 
in unsupervised learning for style transfer of natural images as in \citet{CycleGAN2017}.
Note that paired sets of images at two different domains are not available 
in common, that is the main difficulty in constructing generative 
models for the task of image-to-image translation.
Our training strategy only requires an unpaired set of images at two different domains, 
allowing one to easily increase the number of training data as well as 
examine various translations in an efficient way.

The rest of the present paper is organized as follows.
In Section~\ref{sec:WL}, we summarize the basics of weak gravitational lensing effects. 
In Section~\ref{sec:model}, we describe three generative models of weak lensing mass maps based on multivariate random Gaussian variables, the log-normal model, 
and the generative adversarial networks adopted in this paper.
We provide how to produce the data set for training and 
testing the networks in Section~\ref{sec:data}. 
In Section~\ref{sec:result}, 
we show the performance of our trained networks at the sky coverage as same as in the training datasets. 
We then study the applicability of our method in extending a field of view 
in Section~\ref{sec:extension_sky}. 
Limitations of our generators are listed in Section~\ref{sec:limitation}.
Concluding remarks and discussions are given in Section~\ref{sec:conclusion}.
 
\section{Weak Gravitational Lensing}\label{sec:WL}

The image distortion of a source object (galaxy) induced by weak gravitational lensing is commonly characterized by the $2\times2$ matrix below:
\beqa
A_{ij} = \frac{\partial \theta_{\rm true}^{i}}{\partial \theta_{\rm obs}^{j}}
           \equiv \left(
\begin{array}{cc}
1-\kappa -\gamma_{1} & -\gamma_{2}-\omega  \\
-\gamma_{2}+\omega & 1-\kappa+\gamma_{1} \\
\end{array}
\right), \label{eq:distortion_tensor}
\eeqa
where 
$\kappa$ is the convergence, $\gamma$ is the shear, and $\omega$ is the rotation, 
$\bd{\theta}_{\rm obs}$ and $\bd{\theta}_{\rm true}$ represent the observed position of the source object and the true (unlensed) position, respectively.
In the limit of $\kappa, \gamma \ll 1$ which we are interested in, 
we can express the convergence as the integral of the density contrast of underlying matter density field $\delta_{\rm m}(\bd{x})$ with a weight function over redshift \citep{Bartelmann:1999yn},
\beqa
\kappa(\bd{\theta}) &=& \int_{0}^{\infty}{\rm d}\chi\, W_{\kappa}(\chi) \delta_{\rm m}(r(\chi)\bd{\theta}, \chi), \\ \label{eq:delta2kappa}
W_{\kappa}(\chi) &=& \frac{3}{2}\left(\frac{H_{0}}{c}\right)^2\Omega_{\rm m0}(1+z(\chi))r(\chi) \nonumber \\
&&
\,\,\,\,\,
\,\,\,\,\,
\,\,\,\,\,
\,\,\,\,\,
\times
\int_{\chi}^{\infty}{\rm d}\chi^{\prime}\,
p(\chi^{\prime})\frac{r(\chi^{\prime}-\chi)}{r(\chi^{\prime})},
\label{eq:lens_kernel}
\eeqa
where 
$c$ is the speed of light,
$H_{0}$ is the present-day Hubble constant,
$\Omega_{\rm m0}$ is the matter density parameter at present, 
$\chi(z)$ is the radial comoving distance to redshift $z$,
$r(\chi)$ is the angular diameter distance, and $p(\chi)$ represents 
the source distribution normalized to $\int{\rm d}\chi\,p(\chi)=1$.
Throughout this paper, we assume a single source plane at redshift of $z_{\rm source} = 1$, i.e. $p(\chi)=\delta_\mathrm{D}(\chi-\chi_{1})$ where $\delta_\mathrm{D}(x)$ is the Dirac delta function and $\chi_1=\chi(z=1)$ for simplicity\footnote{We also examined two different source redshifts of $z_\mathrm{source}=0.5$ and $1.5$ to train our neural-network-based model. We found that the model shows a similar performance as in Section~\ref{sec:result} even for $z_\mathrm{source}=0.5$ and $1.5$
as long as we trained the model with labeling of training datasets based on field variances of lensing convergence fields. See Section~\ref{subsec:training} for the labeling.}.

\section{Model}\label{sec:model}

\subsection{Gaussian}\label{subsec:gaussian}

The simplest generative model of the lensing convergence field is 
given by a multivariate random Gaussian distribution. 
The full statistical information of $\kappa$ can be characterized by 
the distribution in Fourier mode $\tilde{\kappa}_{\bm \ell}$.
In general, the distribution of Fourier mode $\kappa_{\bm \ell}$ 
is expressed as the function of norm $|\tilde{\kappa}_{\bm \ell}|$
and phase $\theta_{\kappa, {\bm \ell}}$.
Assuming zero mean, we can write the distribution of Gaussian $\kappa_{\bm \ell}$ as
\beqa
{\rm Prob}(|\tilde{\kappa}_{\bm \ell}|, \theta_{\kappa, {\bm \ell}})
&\equiv& {\cal P}_{G}(|\tilde{\kappa}_{\bm \ell}|) {\cal P}_{G}(\theta_{\kappa, \bm \ell}), \label{eq:gauss_fourier} \\
{\cal P}_{G}(|\tilde{\kappa}_{\bm \ell}|) 
&=& 
\frac{2|\tilde{\kappa}_{\bm \ell}|}{[C(\ell)]^{3/2}}
\exp\left[-\left(\frac{|\tilde{\kappa}_{\bm \ell}|}{\sqrt{C(\ell)}}\right)^2\right], \label{eq:gauss_fourier_norm} \\
{\cal P}_{G}(\theta_{\kappa, {\bm \ell}}) 
&=& 
\frac{1}{2\pi}, \label{eq:gauss_fourier_phase}
\eeqa
where $C(\ell)$ is the power spectrum of $\kappa$ and it is defined by
\beqa
\langle \tilde{\kappa}(\bm{\ell}_1) \tilde{\kappa}(\bm{\ell}_2) \rangle 
= (2\pi)^2 \delta_D (\bm{\ell}_{1}+\bm{\ell}_{2})C (\ell_1).
\eeqa

\ms{The Gaussian generative model has an advantage of being computational efficient. It enables us to produce a map with an arbitral angular resolution and sky coverage. In the concordance cosmological model, the initial condition of density perturbations follows a random Gaussian variable \citep[e.g.][]{2020A&A...641A...9P}, allowing the Gaussian model to be the zeroth order approximation. Nevertheless, non-linear gravitational growth of matter density fields through cosmic ages causes the couping among different Fourier modes of the initial density perturbations \cite[e.g.][for a review]{2002PhR...367....1B}, leading to the emergence of non-Gaussian structures in lensing converegence fields. The Gaussian generative model should remain valid for large-scale Fourier modes, because the large-scale modes are less affected by the gravitational mode coupling.}

\subsection{Log Normal}

Non-linear gravitational growth in the matter density contrast $\delta_\mathrm{m}$
makes the lensing convergence non-Gaussian.
A prominent non-Gaussian feature of $\kappa$ in numerical simulations 
is the presence of a heavy positive tail 
in one-point probability distribution function (PDF) of $\kappa$ \citep[e.g.][]{Jain:1999ir, Takahashi:2011qd, Castro:2017tbn}.
An empirical log-normal model can account for the tail in a simple way \citep{Taruya:2002vy}.
The model assumes that the real-space convergence field follows the one-point PDF of
\beqa
\mathrm{Prob}(\kappa) &=& \frac{1}{\sqrt{2\pi} \sigma_{\ln}}\frac{1}{\kappa+|\kappa_\mathrm{min}|} \nonumber \\
&&
\times \exp\left\{-\frac{\left[\ln(1+\kappa/|\kappa_\mathrm{min}|)+\sigma^2_{\ln}/2\right]^2}{2\sigma^2_{\ln}}\right\}, \label{eq:lognormal_pdf}
\eeqa
for $\kappa > \kappa_\mathrm{min}$, and it holds that $\mathrm{Prob}(\kappa)=0$ otherwise.
In Eq.~(\ref{eq:lognormal_pdf}), the variance $\sigma^2_{\ln}$ is given by 
\beqa
\sigma^2_{\ln} \equiv \ln \left(1+\frac{\langle \kappa^2 \rangle}{|\kappa_\mathrm{min}|^2}\right),
\eeqa
where $\langle \kappa^2 \rangle$ is the variance of $\kappa$. 
Note that the variance is determined by the convergence power spectrum of $C$.

One can rewrite the log-normal model in terms of a new Gaussian variable $y$
by considering a local-type translation of 
\beqa
y(\bd{\theta}) = \ln\left(1+\frac{\kappa(\bd{\theta})}{|\kappa_\mathrm{min}|}\right)+\frac{\sigma^2_{\ln}}{2}. \label{eq:lognormal2Gaussian}
\eeqa
This new Gaussian field $y$ has zero mean and unit variance.
Using the relation of Eq.~(\ref{eq:lognormal2Gaussian}) and the PDF of Eq.~(\ref{eq:lognormal_pdf}), one expect that the two-point correlation of $y$ should be related with the counterpart of $\kappa$ as
\beqa
\xi_{y}(\theta) &\equiv& \langle y(\bd{\theta}+\bd{\phi}) y(\bd{\phi}) \rangle, \\
\xi_{\kappa}(\theta) &\equiv& \langle \kappa(\bd{\theta}+\bd{\phi}) \kappa(\bd{\phi}) \rangle \nonumber \\
&=& \frac{1}{2\pi[1-\xi^2_y(\theta)]^{1/2}}\int_{-\infty}^{\infty}\mathrm{d}u_1\int_{-\infty}^{\infty}\mathrm{d}u_2 \nonumber \\
&&
\times \exp \left(-\frac{u^2_1}{2[1-\xi^2_y(\theta)]}\right)\exp\left(-\frac{u^2_2}{2}\right) \nonumber \\
&&
\times {\cal F}(u_1+\xi_y(\theta)u_2){\cal F}(u_2), \label{eq:kappa_2pcf_lognormal}
\eeqa
where ${\cal F}$ is the inverse of Eq.~(\ref{eq:lognormal2Gaussian}), i.e.~${\cal F}(u) = |\kappa_\mathrm{min}|\exp(u \sigma_{\ln} - \sigma^2_{\ln}/2)-|\kappa_\mathrm{min}|$.
Note that the two-point correlation of $\xi_{\kappa}(\theta)$ can be computed as
the Fourier transform of the power spectrum $C$;
\beqa
\xi_{\kappa}(\theta) = \int\frac{\mathrm{d}^2\ell}{(2\pi)^2}\, C(|\bd{\ell}|) e^{-i\bd{\ell}\cdot\bd{\theta}}.
\eeqa
Therefore, Eq.~(\ref{eq:kappa_2pcf_lognormal}) sets $\xi_{y}$ for a given convergence power spectrum. In the end, the log-normal generative model is fully specified by two parameters of the minimum convergence $\kappa_\mathrm{min}$ and convergence power spectrum $C$.

\subsection{Cycle-Consistent Generative Adversarial Networks}

The log-normal model partly accounts for non-linear gravitational effects on the lensing convergence, while its validity is still limited. For instance, the log-normal model cannot reproduce summary statistics sensitive to rare events such as skewness and kurtosis \citep{Taruya:2002vy}. Also, multiple-point correlation functions in the log-normal model can be expressed as a function of the two-point counterpart \citep{Hilbert:2011xq}, but this is not the case in numerical simulations of cosmic large-scale structures \citep{Colavincenzo:2018cgf, Piras:2022dgt}.

To improve the log-normal model, we here examine a machine-learning approach for image synthesis. The GAN is a generative model \citep{NIPS2014_5ca3e9b1} that contains two neural networks: a generator $G$ and a discriminator $D$. Given the real data set, $G$ aims to create false\footnote{\ms{We mean that the ``false" data are the data produced through a different generation process from the real counterpart.}} data that looks like the genuine data from the real sample, while $D$ tries to discriminate the false data and the genuine counterpart. The GAN can be effectively trained with the back-propagation algorithm and appropriate network architecture even when limited training data are available 
\citep[e.g. see][for a review]{2018arXiv180304469H}.

Among several models of the GAN, we adopt the Cycle-Consistent GAN (Cycle GAN) consisting of a pair of GANs \citep{CycleGAN2017}.
The Cycle GAN is designed to learn a style transfer between two different image domains.
A representative example is the translation between 
a landscape photograph and a Monet painting. 
Because there are no paired image sets between landscape photographs and Monet paintings in practice, one requires the Cycle GAN to train with unpaired image collections, consisting of a source set of $\{x_i\}_{i=1}^{N}\, (x_i \in \mathrm{landscape \, photos})$ and a target set $\{y_j\}_{j=1}^{M}\, (y_j \in \mathrm{Monet\, paintings})$.
In this setup, one is not able to use the information that $x_i$ in the input domain matches to $y_j$ in the target for training of the Cycle GAN.
In this paper, we train the Cycle GAN with the network architecture proposed in \citet{CycleGAN2017} by using unpaired image sets of the Gaussian convergence and the convergence produced in lensing simulations of \citet{Liu:2017now}.
It would be worth noting that the Gaussian convergence model is computationally efficient, while a massive production of lensing simulations requires high computational costs.
This situation is similar to the translation between a landscape photograph and a Monet painting, i.e. it is easy to prepare a large set of landscape photographs with low costs, but it is really expensive (impossible in fact) to increase the number of Monet paintings.

Details of the unpaired sets and training process are described in the next section, while the network architecture in the Cycle GAN is provided in Appendix~\ref{apdx:network}.
\ms{We use the Pytorch implementation\footnote{\url{https://github.com/junyanz/pytorch-CycleGAN-and-pix2pix}} of the Cycle GAN throughout this paper. Note that the network architecture is same as developed in \citet{CycleGAN2017}. The architecture has been examined for various image-to-image translations in real-world examples.}

We here briefly summarize how to train the Cycle GAN in an unsupervised fashion.
Let $X$ and $Y$ be the image sets produced by the Gaussian model as in Section~\ref{subsec:gaussian} and the set of the lensing convergence in the simulations, respectively.
We refer to $G_X$ as the generator translating an input image 
of $x \in X$ to $y \in Y$, whereas $D_X$ 
judges if an input image is produced by $G_X$ or not.
We also define the generator of $G_Y$ and the discriminator of $D_Y$ for the translation of $y \rightarrow x$ in a similar way.
Our objective is then summarized as
\beqa
G_{X}, G_{Y} = \mathrm{arg}\min_{G_X, G_Y} \max_{D_X, D_Y}{\cal L}(G_X, G_Y, D_X, D_Y),
\eeqa
where ${\cal L}(G_X, G_Y, D_X, D_Y)$ is the loss function for the Cycle GAN 
containing four parts below;
\beqa
{\cal L}(G_X, G_Y, D_X, D_Y) &=& 
{\cal L}_\mathrm{GAN}(G_X, D_X) + {\cal L}_\mathrm{GAN}(G_Y, D_Y) \nonumber \\
&& 
\quad \quad \quad \quad
+ \lambda_1 {\cal L}_\mathrm{cyc}(G_X, G_Y) \nonumber \\
&&
\quad \quad \quad \quad
+ \lambda_2 {\cal L}_\mathrm{id}(G_X, G_Y). \label{eq:cycle_GAN_loss}
\eeqa
The first two functions in the right-hand side of Eq.~(\ref{eq:cycle_GAN_loss})
represent the losses for the two GANs and those are defined by
\beqa
{\cal L}_\mathrm{GAN}(G_X, D_X) &=& \mathbb{E}_{y} \log\left[D_X(y)\right] \nonumber \\
&&
+ \mathbb{E}_{x} \log\left\{1-D_X\left[G_X(x)\right]\right\}, \label{eq:GAN_loss_X}\\
{\cal L}_\mathrm{GAN}(G_Y, D_Y) &=& \mathbb{E}_{x} \log\left[D_Y(x)\right] \nonumber \\
&&
+ \mathbb{E}_{y} \log\left\{1-D_Y\left[G_Y(y)\right]\right\}, \label{eq:GAN_loss_Y}
\eeqa
where $\mathbb{E}_x$ represents the expected value over a sample of $x$ and so on.
The term of ${\cal L}_\mathrm{cyc}$ in Eq.~(\ref{eq:cycle_GAN_loss}) regularizes the two generators of $G_X$ and $G_Y$ so that the learned mapping between $X$ and $Y$ should be cycle-consistent. It is given by
\beqa
{\cal L}_\mathrm{cyc}(G_X, G_Y) &=& \mathbb{E}_{x}\left|G_Y\left[G_X(x)\right]-x\right|_\mathrm{1} \nonumber \\
&&
+ \mathbb{E}_{y}\left|G_X\left[G_Y(y)\right]-y\right|_\mathrm{1},
\eeqa
where $|\cdots|_\mathrm{1}$ is the L1 norm.
We further constrain the generators with the term of ${\cal L}_\mathrm{id}(G_X, G_Y)$
so that they can provide an identity mapping when real samples of the target domain are used as their input, i.e. 
\beqa
{\cal L}_\mathrm{id}(G_X, G_Y) &=& \mathbb{E}_{y}\left|G_X(y)-y\right|_\mathrm{1} 
+ \mathbb{E}_{x}\left|G_Y(x)-x\right|_\mathrm{1}.
\eeqa
We have two hyperparameters of $\lambda_1$ and $\lambda_2$ in the loss and 
set $\lambda_1=10$ and $\lambda_2 = 5$ throughout this paper\footnote{As a sanity check, we examined different hyperparameters of $(\lambda_1, \lambda_2) = (20, 5), (5,5), (10, 2.5)$ and $(10, 10)$. We found the results in Section~\ref{sec:result} are minimally affected by the change of each hyperparameter by a factor of 2 or 0.5.}.
\ms{In our setup, the reguralization terms are typically 10 times as small as other GAN losses. Because the relative differences between the GAN losses and the regularization terms are important, we add $\lambda_1$ and $\lambda_2$ to the ${\cal L}_\mathrm{cyc}$ and ${\cal L}_\mathrm{id}$ terms alone.}

\section{Data and Methods}\label{sec:data}
\subsection{High-cost Lensing Simulations}

To train the Cycle GAN, we use publicly available numerical simulations of weak gravitational lensing as performed in \citet{Liu:2017now}.
The simulation suite is referred to as MassiveNuS (Cosmological Massive Neutrino Simulations), aiming at detailed studies of cosmic large-scale structures in the presence of massive neutrinos.
MassiveNuS consists of 101 models in flat-$\Lambda$CDM cosmology.
In this paper, we adopt the simplest flat-$\Lambda$CDM model 
including massless neutrinos with cosmological parameters below; 
total mass density $\Omega_\mathrm{m0}=0.3$, 
baryon density $\Omega_\mathrm{b0}=0.046$,
cosmological constant $\Omega_\Lambda=1-\Omega_\mathrm{m0}=0.7$,
present-day Hubble parameter $H_0 = 100h = 70.0\, \mathrm{km}/\mathrm{s}/\mathrm{Mpc}$, 
primordial scalar spectrum power-law index $n_s=0.97$,
and the primordial curvature power spectrum at the pivot scale $k=0.05 \, \mathrm{Mpc}^{-1}$ being $A_s = 2.1\times10^{-9}$.
The authors in \citet{Liu:2017now} run N-body simulations
with the public tree-Particle Mesh code Gadget-2 \citep{Springel:2005mi}, 
adopting a box size of $512\, h^{-1}\mathrm{Mpc}$ and $1024^3$ particles. 
The Zel'dovich approximation \citep{Zeldovich:1969sb} was adopted 
when they generated the initial condition at $z=99$ for the N-body simulations.
From the N-body simulation snapshots, they then produced 10,000 convergence maps at five different source redshifts of $z_\mathrm{source} = 0.5, 1.0, 1.5, 2.0$, and $2.5$ by using the public ray-tracing code LensTools \citep{Petri:2016zrq}.
Each map covers a sky of $3.5^2\, \mathrm{deg}^2$ with the number of pixels on a side being 512.
In this paper, we degrade the convergence map at $z_\mathrm{source}=1.0$ 
from the original resolution of $512\times512$ to $256\times256$ as it is used for the input of the generators in our GAN. 

\subsection{Low-cost Gaussian Simulations}

As another training dataset for the Cycle GAN, we produce 10,000 convergence maps assuming the Gaussian model as described in Section~\ref{subsec:gaussian}.
In the Gaussian model, we set the convergence power spectrum of $C$ so that 
it can be in agreement with the average power spectrum in the lensing simulations.
To be specific, we assume a functional form of the power spectrum below;
\beqa
C(\ell) = C_\mathrm{ref}(\ell)\, \left[1+\left(\frac{\ell}{\ell_0}\right)^{\alpha}\right]^{-\beta}, \label{eq:Cell_massivenus}
\eeqa
where $C_\mathrm{ref}(\ell)$ represents a reference model of the power spectrum,
and there are three free parameters of $\ell_0$, $\alpha$ and $\beta$ to account for resolution effects in the lensing simulations.
Using Eq.~(\ref{eq:delta2kappa}) with the Limber approximation \citep{Limber:1954zz},
we compute the reference model of $C_\mathrm{ref}$ as
\beqa
C_\mathrm{ref}(\ell) = \int_{0}^{\infty} {\rm d}\chi \frac{W^2_\kappa(\chi)}{r^2(\chi)} 
P_{\delta}\left(k=\frac{\ell}{r(\chi)},z(\chi)\right)
\label{eq:kappa_power},
\eeqa
where $P_{\delta}(k,z)$ is the three dimensional matter power spectrum at redshift $z$,
and it is computed as the fitting formula developed in \citet{Takahashi:2012em}.
We constrain the free parameters of $\ell_0$, $\alpha$, and $\beta$ by least-squares fitting for the power spectrum in the lensing simulations. 
When estimating the convergence power spectrum from the simulations,
we first adopt the fast Fourier transform of the convergence field and 
then measure the binned power spectrum of the convergence field 
by averaging the product of Fourier modes $|\kappa_{\bm \ell}|^2$.
We employ 15 bins logarithmically spaced in the range of $\ell=100$ to $\ell=10^4$.
In the end, we find that Eq.~(\ref{eq:kappa_power}) can explain the average power spectrum over 10,000 simulations at $z_\mathrm{source}=1$ within a $5\%$-level accuracy when setting $\ell_0=1.360\times10^{4}$, $\alpha=1.643$, and $\beta=4.088$.
This pre-training of the Gaussian model allows the Cycle GAN to learn non-Gaussian information about the lensing simulations in an efficient way.
Also, it is important to prepare training sets with reasonable similarities when one trains the GAN for an image-to-image translation \citep{8100115, CycleGAN2017, Shirasaki:2018thk}.

\subsection{Training Cycle GAN}\label{subsec:training}

\begin{figure}[!t]
\includegraphics[clip, width=1.\columnwidth]{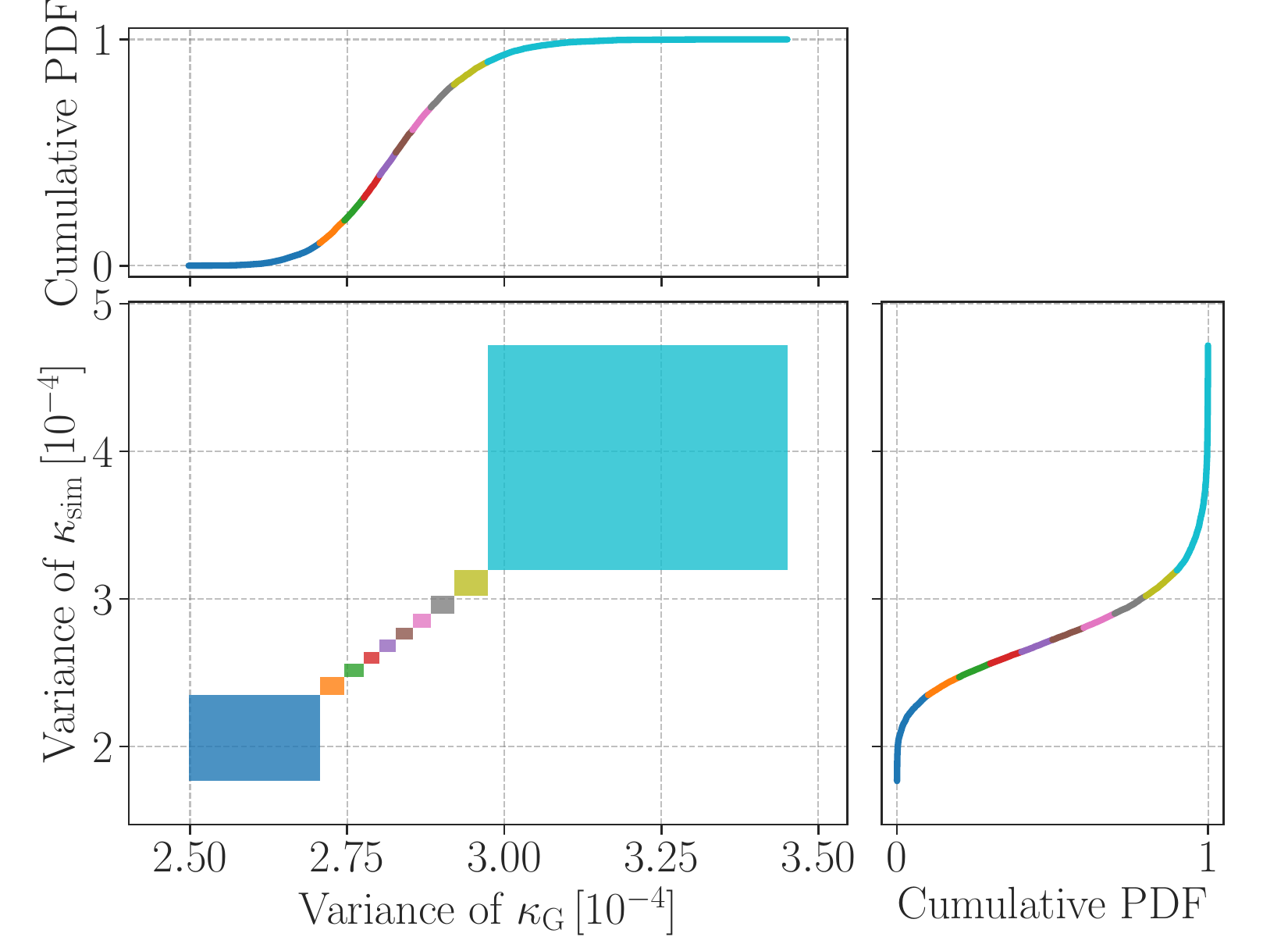}
\caption{\label{fig:var_labels} 
Labeling of training datasets with field variances of lensing convergence fields. Different colored squares in the lower left panel stand for labels of two different datasets. The labels are set by the field variance of $\langle \kappa^2 \rangle$ for either simulated or Gaussian fields. 
\ms{The size of each square represents a width of field variances. 
Each square contains the same number of realizations of training data.}
In this figure, we consider 10 labels. 
The horizontal axis in the lower left represents the variance of the Gaussian convergence fields, while the vertical axis shows the simulated counterpart. The upper and right panels show the cumulative distribution of the variances over 10,000 realizations of the two different fields. 
When training the Cycle GAN, we use the datasets with identical labels, 
requiring each of 10 different generators to be trained with 1000 realizations of the training sets. It is worth noting that each realization of Gaussian and simulated fields is not aligned, e.g. the two do not share phase information in a survey window with each other.
}
\end{figure}

For training the Cycle GAN, we follow the procedure as adopted in \citet{CycleGAN2017}.
To make the training stable, we replace the negative log-likelihood objectives in Eqs.~(\ref{eq:GAN_loss_X}) and (\ref{eq:GAN_loss_Y}) 
with least-squares loss functions \citep{8237566}.
In practice, we train $G_X$ to minimize $\mathbb{E}_x[D_X(G_X(x))-1]^2$
and then update the parameters in $D_X$ to minimize 
$\mathbb{E}_y[D_X(y)-1]^2 + \mathbb{E}_x[D_X(G_X(x))]^2$
in Eq.~(\ref{eq:GAN_loss_X}) at every batch.
A similar replacement is applied in Eq.~(\ref{eq:GAN_loss_Y}).
Furthermore, we update the discriminators 
using a history of generated images rather than the
ones produced by the latest generators. 
We keep an image buffer that stores the 50 previously created images.
This strategy is effective in reducing a gap between synthetic and 
real image distributions \citep{8099724}.
We use the Adam optimizer \citep{KingBa15} with a batch size of 1 to update the parameters in the Cycle GAN. The batch size of 1 is known to provide better results when the generator has a U-net architecture as we adopted \citep{8100115}. 
All networks are trained with a learning rate of 0.0002 for the first 100 epochs, while we linearly decay the rate to zero over the next 100 epochs. 
Weights are initialized from a Gaussian distribution with zero mean and standard deviation of 0.02.

\begin{table*}[t]
\caption{\label{tab:all_summary_stats}%
Summary statistics of lensing convergence fields. Those statistics are used for validation of the performance of our neural network-based generative model.
We are interested in not only the ensemble average but also the covariance matrix of each statistic.
}
\begin{ruledtabular}
\begin{tabular}{lll}
Name &
Symbol &
Note \\
\colrule
Power Spectrum & $C$ & Variance of Fourier modes at a given multipole $\ell$ \\
Bispectrum & $B$ & Three-point correlations among Fourier modes at a given triangle configuration \\
Peak counts & $N_\mathrm{pk}$ & Number density of local maxima \\
One-point PDF & ${\cal P}$ & Histogram of pixel values \\
Minkowski functionals (MFs) & $V_1$ and $V_2$ & Total boundary length and integral of
geodesic curvature along contours \\
Scattering transform coefficients & $s_1$ and $s_2$ & A compressed set of two- and four-point spatial correlations
\end{tabular}
\end{ruledtabular}
\end{table*}

When using the training datasets without any annotations, we find that convergence fields produced by the generator $G_X$ can reproduce some summary statistics observed in the lensing simulations on average, whereas $G_X$ fails to predict the variance as in the lensing simulations. 
To improve the performance of the generator, we divide the training sets into $N$ groups by using a label that is relevant to variations in the convergence fields.
Among several candidates, we decided to use the field variance 
$\langle \kappa^2\rangle$ to measure the variations. 
This choice is physically motivated as the scatter in the lensing power spectrum $C$ at large $\ell$ is mostly determined by four-point correlations with squeezed quadrilaterals including a shared infinite wavelength mode \citep{Takada:2013wfa}.
According to a halo-model prescription \citep{Cooray:2002dia}, the large-$\ell$ variance in $C$ depends on the variance of matter density in a survey window, sharing some information with the field variance of $\langle \kappa^2 \rangle$.
Hence, we expect that the field variance $\langle \kappa^2 \rangle$ provides a good measure of the variation in lensing convergence fields.
An example of labeling the training sets is shown in Figure~\ref{fig:var_labels}.
In the figure, we divide the training sets into $N=10$ groups by their field variances.
We then train the Cycle GAN by using the lensing simulations and the Gaussian data with the same label of $i$ ($i=1, 2, \cdots, N$).
In the end, we will have $N$ different generators 
to produce non-Gaussian convergence fields from given Gaussian fields. 
When producing a new non-Gaussian field, 
we select which generator will be used by the field variance of an input Gaussian field. 
Throughout this paper, we set $N=10$ for our fiducial choice.
The $N$ dependence of our results is examined in Section~\ref{subsec:results_powerspec}.

As a by-product, the Cycle GAN provides a generator to translate non-Gaussian convergence fields into Gaussian counterparts. Although this translation is 
not our primary purpose, it is interesting to see 
if the Cycle GAN can be used for a reasonable Gaussianization of weak lensing mass maps.
Note that appropriate Gaussianization enhances the cosmological constraining power of weak lensing fields \citep{YuYu:2011ijg,2011ApJ...729L..11S,Seo:2011ku,Simpson:2015nva}.
Appedix~\ref{apdx:RT2Gauss} summarizes the results of the translation from the non-Gaussian to the Gaussian mass maps. 
A simple training without any labels can gaussianize the non-Gaussian weak lensing mass maps obtained by the ray-tracing simulation with a reasonable precision;
The Gaussianized power spectrum can reproduce the true Gaussian counterpart with a level of $2\%$ at $\ell \simgt 400$, while the Gaussianized one-point PDF is well described by the expected Gaussian distribution. Also, the average bispectrum over  10,000 Gaussianized lensing mass maps is consistent with a null detection within the statistical uncertainty for a sky coverage of $3.5^2\, \mathrm{deg}^2$.
We also find that field-variance-based labels do not help to improve the performance in the translation when the target domain is set to the Gaussian model.

\subsection{Summary Staistics}\label{subsec:summary_stats}

Because our neural-network-based model is trained in an unsupervised fashion,
it is non-trivial to assess the performance of our model.
In fact, the loss for GANs does not provide any measures for model validation. 
In this paper, we validate the model by studying various summary statistics of generated (fake) convergence fields and comparing them with their counterparts 
in the ground-truth lensing simulations.
Because of non-Gaussian natures in the lensing convergence, 
several summary statistics have been studied in the literature.
Apart from the standard power spectrum, 
we here consider five non-Gaussian statistics of 
bispectrum, peak counts, one-point probability distribution function (PDF), 
Minkowski functionals (MFs), and scattering transform coefficients.
Characteristics of each non-Gaussian statistic are briefly summarized below.
Table~\ref{tab:all_summary_stats} provides all the summary statistics of interest in this paper with a short description.

\subsubsection*{Bispectrum}

The bispectrum is the three-point correlation in Fourier space and is defined as
\beqa
\langle \tilde{\kappa}(\bm{\ell}_1) \tilde{\kappa}(\bm{\ell}_2) \tilde{\kappa}(\bm{\ell}_3)\rangle 
&=& (2\pi)^2 \delta_D (\bm{\ell}_{1}+\bm{\ell}_{2}+\bm{\ell}_{3}) \nonumber \\
&&
\quad \quad \quad \quad
\times B (\bm{\ell}_{1}, \bm{\ell}_{2}, \bm{\ell}_{3}),
\eeqa
where $B$ represents the bispectrum.
Because it holds that $B=0$ at any Fourier-mode configurations for the Gaussian convergence model, the bispectrum provides a powerful means of 
studying the lowest-order non-Gaussian information in the convergence field \citep{Hui:1999ak, Cooray:2000uu, Benabed:2001dm, Takada:2003ef, Refregier:2003xe, Dodelson:2005rf, Coulton:2018ebd}.
As in the case of power spectra, one can relate the bispectrum to 
the three-dimensional matter bispectrum $B_{\delta}$;
\beqa
B(\bm{\ell}_{1}, \bm{\ell}_{2}, \bm{\ell}_{3})
&=& \int_{0}^{\infty} {\rm d}\chi \frac{W^3_\kappa(\chi)}{r^4(\chi)} \nonumber \\
&&
\quad \quad 
\times B_{\delta}\left(\bm{k}_{1}, \bm{k}_{2}, \bm{k}_{3}, z(\chi)\right),
\label{eq:kappa_bispec}
\eeqa
where we evaluate the wavenumbers in the three-dimensional space as ${\bm{k}_{i} = \bm{\ell}_{i}/r(\chi)}$. 
Theoretical and numerical studies have predicted that 
the bispectrum can add a $20-50$ percent-level gain 
to signal-to-noise ratio of the power spectrum 
up to a maximum multipole of a few thousands 
\citep{2013MNRAS.429..344K, Sato:2013mq}.

To estimate the bispectrum from a given pixelized data of the convergence, 
we measure an average of the product of three Fourier modes 
$\mathrm{Re}[\tilde{\kappa}(\bm{\ell}_{1})\tilde{\kappa}(\bm{\ell}_{2})\tilde{\kappa}(\bm{\ell}_{3})]$ 
over a triangle configuration of $\bm{\ell}_{1}+\bm{\ell}_{2}+\bm{\ell}_{3}=\bm{0}$
where $\mathrm{Re}[\cdots]$ stands for the real part of a complex number. 
Throughout this paper, 
we adopt 15 bins logarithmically spaced in the range of 
$\ell_i \, (i = 1, 2, 3) = 100$ to $10^{4}$ for each of the three multipoles.

\subsubsection*{Peak Counts}

The number count of local maxima in a convergence map referred to as peak counts, has
become a popular summary statistic to extract non-Gaussian cosmological information beyond the standard power-spectrum-based analysis in practice \citep{Liu:2014fzc,
Hamana:2015bwa, DES:2016jfa, Shan:2017mgz, Martinet:2017rqp, Harnois-Deraps:2020pvj, DES:2021epj, Liu:2022gnc}.
Peaks with heights greater than a $\sim4\sigma$ noise level 
arise from single massive dark matter haloes \citep{Hamana:2003ts, Hennawi:2004ai, Maturi:2004rn, 2012MNRAS.423.1711M, 2012MNRAS.425.2287H}, whereas peaks with modest heights can be created by superposition of large-scale structures and multiple dark matter halos \citep{2011PhRvD..84d3529Y, Liu:2016xjb, Sabyr:2021vpr}.

In this paper, we define the peak as a pixel that has a higher value than all of its eight neighbor pixels in the discretized convergence field. Note that we do not apply any additional smoothing to count peaks. We consider 41 bins in the range of $-0.05\le \kappa \le 0.15$. Since a typical field variance of the convergence field in our simulations is found to be $3\times10^{-4}$ (see Figure~\ref{fig:var_labels}), the largest peak heights in our analysis correspond to a $\sim9\sigma$ significance level. 

\subsubsection*{One-point Probability Distribution Function}

The one-point PDF of the convergence contains vital information 
about the non-Gaussian matter density field \citep{Bernardeau:2000et, Liu:2018dsw, Barthelemy:2020yva, Thiele:2020rig, Boyle:2020bqn, Giblin:2022ucn}.
It encodes the information originating from moments of all orders in the convergence, 
allowing us to study the non-Gaussian nature of the weak lensing field 
at a given length scale.
As a one-point statistic, the PDF can be easily obtained from simulated 
and real data even in the presence of complex survey boundaries \citep{DES:2017hhj, DES:2017eav}. 
In this paper, we measure the PDF from a histogram using 41 linearly spaced bins in the range of $-0.05\le \kappa \le 0.15$. 

\subsubsection*{Minkowski Functionals}

MFs are a useful statistic to study morphological information encoded in smoothed random fields. For two-dimensional maps as the lensing convergence, three MFs of 
$V_0$, $V_1$, and $V_2$ are given by the area in which the convergence is above a threshold of $\kappa_\mathrm{thre}$, 
the total boundary length, and the integral of geodesic curvature along the contours, respectively.
In particular, 
$V_2$ is equal to the number of connected regions above the threshold, 
minus those below the threshold.
Therefore, as the threshold increases, 
$V_2$ provides almost the same information as the peak counts.

MFs can be evaluated by a perturbative approach when the field of interest exhibits a weak non-Gaussianity \citep{Matsubara:2003yt, 2012MNRAS.419..536M, Matsubara:2020fet},
but it is yet difficult to predict MFs of highly non-Gaussian fields with analytic approaches \citep{Petri:2013ffb}.
Simulation-based predictions of the convergence MFs have been constructed so far,
demonstrating that the MFs can carry cosmological information beyond the standard power spectrum analysis \citep{2012PhRvD..85j3513K, Shirasaki:2013zpa, Petri:2013ffb}.
Note that $V_0$ provides a cumulative PDF of the lensing convergence, 
sharing some information with the aforementioned one-point PDF.
Hence, we focus on the two MFs of $V_1$ and $V_2$ in this paper.
We estimate the MFs from a pixelized map by following the method in \citet{Lim:2011kd, 2012PhRvD..85j3513K}. We compute MFs for 41 equally spaced bins of $\kappa_\mathrm{thre}$ between $-0.05$ and $0.15$. 

\begin{figure*}[!t]
\includegraphics[clip, width=2.1\columnwidth, rviewport= 0.05 0.1 1.0 0.9]
{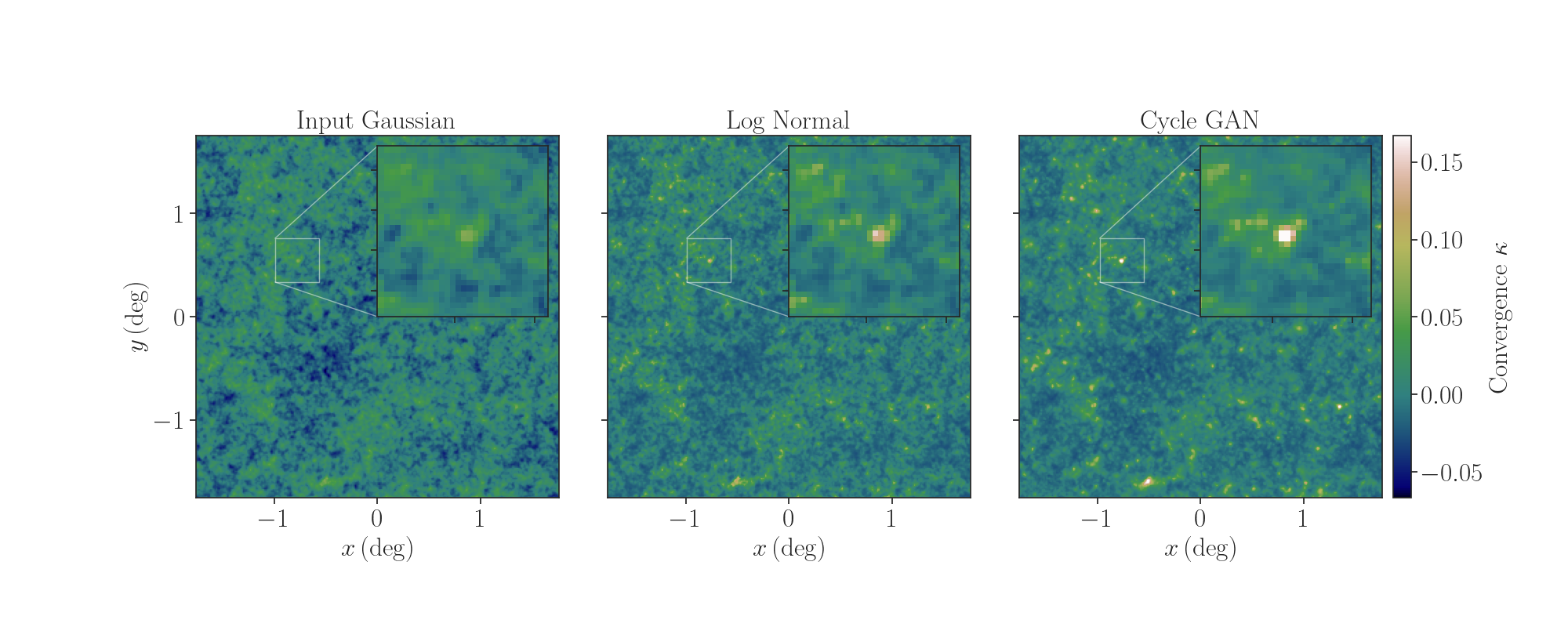}
\caption{\label{fig:cycle_GAN_example} 
An example of lensing convergence fields generated by our neural network.
The left panel shows an input Gaussian field, while the generated convergence by our model is presented on the right. For comparison, the middle panel represents a log-normal model with the same phase information as the Gaussian field.
In each panel, the inset figure highlights a zoom-in view around the highest convergence value in a field of view covering $3.5^2\, \mathrm{deg}^2$.
}
\end{figure*}

\subsubsection*{Scattering Transform Coefficients}

The scattering transform (ST) provides a powerful means of extracting information from high-dimensional data, while it has mathematically tractable properties of translational invariance, non-expanding variance, and Lipschitz continuous to spatial deformation \citep{Mallat2012}. 
The ST generates a set of new fields from a given input field by applying a wavelet convolution and a modulus recursively and uses the expected values of the new fields as
a collection of statistical information encoded in the original input field. 
Hence, the ST performs a similar operation to what a common 
convolutional neural network (CNN) does in practice, 
i.e. either the ST or the CNN relies on the use of localized convolution kernels, and a non-expansive non-linear operator.
\citet{Cheng:2020qbx} has shown that the ST statistic can provide cosmological constraints as tight as the CNN does \citep{Ribli:2019wtw} 
for given weak lensing mass maps.
It would be worth noting that the CNN-based cosmological inference requires a costly training process with simulations, whereas the ST statistic simply needs to emulate its expected values as a function of cosmology (as same as other summary statistics do).
Considering these points, it is interesting to see if fake convergence fields generated by our model can reproduce the ST statistic as in the true lensing simulations.

In this paper, we measure the ST statistic by following the method in \citet{Cheng:2020qbx}.
For a given lensing convergence field $\kappa$, 
we first generate a set of first-order fields as $I_1 \equiv |\kappa \star \Psi^{j_1, l_1}|$, where $\kappa \star \Psi^{j, l}$ represents 
a convolution with wavelets $\Psi^{j,l}$ with a size index $j$ and orientation index $l$,
and $|\cdot|$ stands for the modulus of a complex-value field. 
The second-order fields are then defined as $I_2 \equiv |I_1 \star \Psi^{j_2, l_2}|$.
In this paper, we consider the ST statistic up to second order, which can be interpreted as a compact set of spatial information about two- and four-point correlation functions \citep{Cheng:2020qbx}. See also \citet{Cheng:2021xdw} for a more intuitive understanding of the ST statistic. 

From a set of $\kappa, I_1,$ and $I_2$, we define the ST statistic as
\beqa
S_0 &\equiv& \langle \kappa \rangle, \\
S_1(j_1, l_1) &\equiv& \langle I_1 \rangle, \\
S_2(j_1, j_2, l_1, l_2) &\equiv& \langle I_2 \rangle,
\eeqa
where $\langle \cdot \rangle$ represents the spatial average over a survey window.
We refer to $S_0, S_1$, and $S_2$ as the ST coefficients.
In this paper, we work with Morlet wavelets with $L=4$ different rorations.
One can construct a family of Morlet wavelets by
dilating a prototype wavelet by factors of $2^{j}$ and rotating it by $l \times 45$ degrees.
The prototype Morlet wavelet is defined as 
\beqa
\Psi_0(x, y) &=& \frac{\exp\left(-\frac{x^2+y^2}{2\sigma^2}\right)}{\sqrt{2}\sigma} \nonumber \\
&&
\times \left[\exp\left(ik_0 x\right)-\exp\left(-\frac{k^2_0\sigma^2}{2}\right)\right],
\eeqa
where $\sigma = 0.8\, \mathrm{pixel}$ and $k_0 = 3\pi/4\, \mathrm{pixel}^{-1}$.

As proposed in \citet{Cheng:2020qbx}, we further reduce the ST coefficients by averaging over the rotation indices of $l$ 
because of isotropy in the lensing convergence. 
The relevant ST coefficients to the convergence up to the second order are then given by 
\beqa
s_1(j_1) &=& \frac{\sum_{l_1} S_1(j_1, l_1)}{\sum_{l_1}1}, \\
s_2(j_1, j_2) &=& \frac{\sum_{l_1,l_2} S_2(j_1, j_2, l_1, l_2)}{\sum_{l_1, l_2}1}.
\eeqa
Throughout this paper, we consider the Morlet wavelets with $J=7$ dyadic scales.
In this case, we have 7 and 21 independent elements in $s_1$ and $s_2$, respectively.
We measure the ST coefficients of $s_1$ and $s_2$ from a given convergence field 
with a publicly available code developed by S. Cheng\footnote{\url{https://github.com/SihaoCheng/scattering_transform}}. 

\section{Results}\label{sec:result}

\subsection{Visual impression}

We first see what convergence fields generated by our GAN-based model look like.
Figure~\ref{fig:cycle_GAN_example} shows an example of the fake convergence fields
as well as an input Gaussian convergence and a popular log-normal model.
For the log-normal model, we assume the power spectrum of Eq.~(\ref{eq:Cell_massivenus}) so that the model can reproduce the average power spectrum in the lensing simulations.
We also set the minimum convergence of $\kappa_\mathrm{min}=-0.0367$ for the log-normal model, which is equal to the minimum value over 10,000 lensing simulations.

The right panel in the figure displays the convergence field produced by our generators with the input being the Gaussian field shown on the left.
Our generator adds prominent features at positive convergence regions, 
mimicking highly non-linear gravitational effects in dark matter haloes found in N-body simulations. 
Similar effects can be seen in the log-normal model as shown in the middle, 
but we find that our generators tend to make convergence values higher than the log-normal counterpart. 
This is necessary to make a long tail in the one-point PDF of the fake convergence fields, as predicted by previous numerical simulations \citep{Das:2005yb, Wang:2008hi, Takahashi:2011qd, Shirasaki:2016vve}. 
It is worth noting that our model preserves large-scale phase information encoded in the input Gaussian model.
Hence, the model enables us to produce a set of lensing convergence fields in an intuitive way for cosmological analyses.

\subsection{Power spectrum}\label{subsec:results_powerspec}

\begin{figure*}[!t]
\centering
\includegraphics[clip, width=2.1\columnwidth]{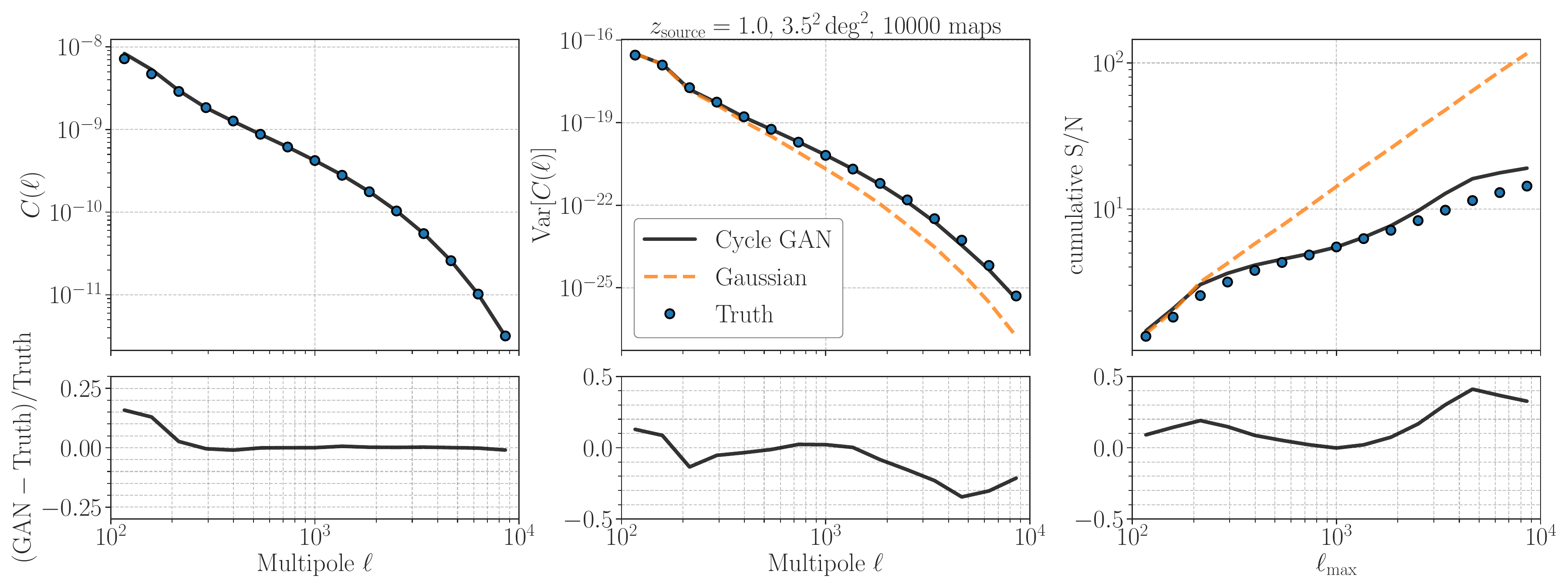}
\caption{\label{fig:comp_powerspec} 
Comparisons of convergence power spectra among several generative models.
The left panel shows the average power spectra, while the middle presents the variance at a given multipole $\ell$. In the right panel, we compare the cumulative signal-to-noise ratio (S/N) defined by Eq.~(\ref{eq:s2n_power}). In each panel, the blue circles represent the results of the true lensing simulations, whereas the black solid and orange dashed lines stand for ones by our GAN-based and the Gaussian models, respectively.
At the bottom, we also show the fractional difference between the GAN predictions and the simulation counterparts, highlighting that our generators can reproduce average, variance, and the cumulative S/N at $300 \simlt \ell \simlt 2000$ within a 10\%-level accuracy.
Note that all the results in this figure are based on 10,000 realizations of the simulations, fake convergence fields generated by our networks, and Gaussian models
at a source redshift being $z_\mathrm{source}=1$. Each realization covers a sky of $3.5^2\, \mathrm{deg}^2$.
}
\end{figure*}

\begin{figure*}[!t]
\includegraphics[clip, width=2.1\columnwidth]{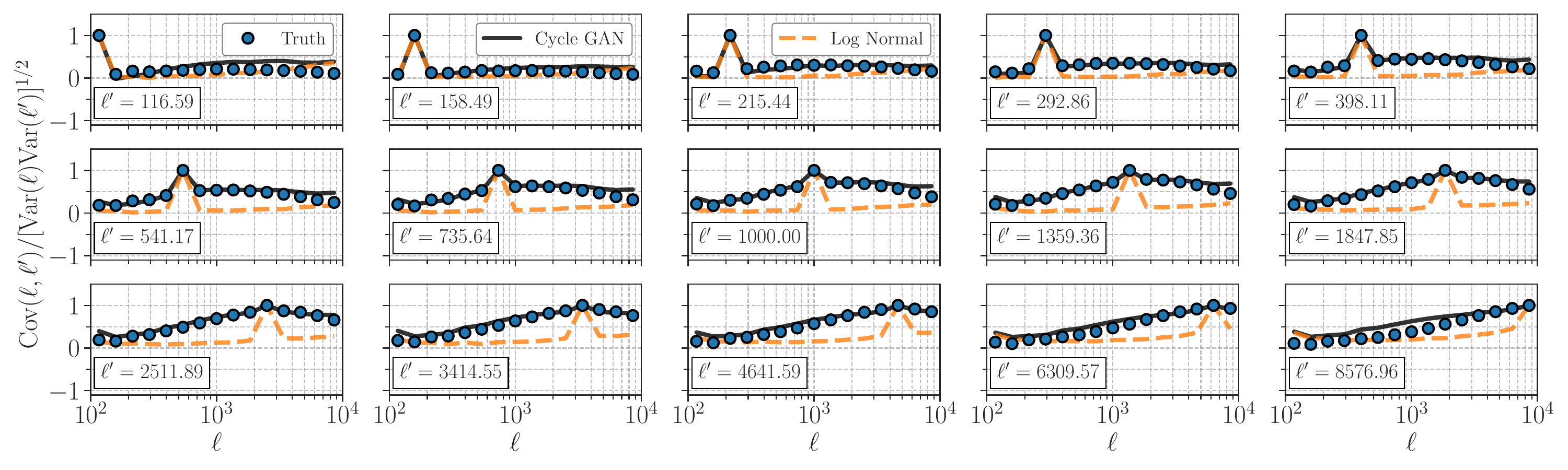}
\caption{\label{fig:comp_powerspec_cov} 
A close look at covariance matrices of convergence power spectra among several models.
Each panel summarizes the covariance at different two scales of $\ell$ and $\ell'$ as normalized to unity at the diagonal elements.
The blue circles in each panel represent the simulation results, while the black solid lines stand for the predictions by our generative model.
For comparison, the orange dashed lines in each panel show the covariance matrix estimated by 10,000 realizations of the log-normal model.
This figure validates that the covariance matrix by our generators can provide a reasonable fit to the simulation results, making our model valuable 
for standard cosmological analyses based on two-point correlation functions.
}
\end{figure*}

\begin{figure*}[!t]
\includegraphics[clip, width=2.1\columnwidth]{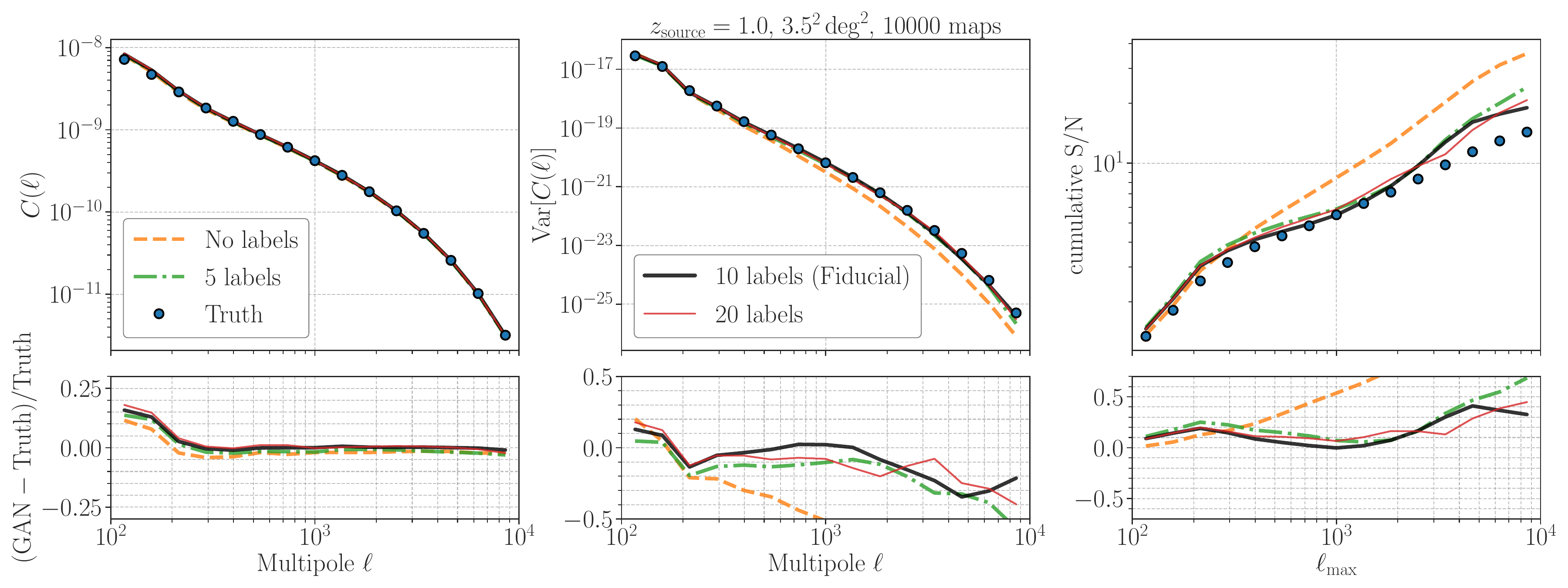}
\caption{\label{fig:comp_powerspec_varying_labels} 
Dependence of our model performance on the number of labels based on field variances.
Similar to Figure~\ref{fig:comp_powerspec}, 
but we show the results when varying the number of labels $N$ in the training datasets.
The blue circles in each panel are the results obtained from 10,000 lensing simulations,
while different colored lines stand for the model predictions with different $N$.
}
\end{figure*}

We then investigate the most basic summary statistic of the convergence power spectrum $C(\ell)$ generated by our neural network-based model.
Figure~\ref{fig:comp_powerspec} summarizes the statistical nature of $C(\ell)$ predicted by our model and compares it with the counterparts of the true lensing simulations.
In the figure, we look into three properties of ensemble average, variance, and a cumulative signal-to-noise ratio defined as
\beqa
\left(\frac{\mathrm{S}}{\mathrm{N}}\right)[\ell_\mathrm{max}] = \left[\sum_{\ell_i, \ell_j \le \ell_\mathrm{max}}C(\ell_i) \, \mathrm{Cov}^{-1}(\ell_i, \ell_j) \, C(\ell_j)\right]^{1/2},
\label{eq:s2n_power}
\eeqa
where $C(\ell_i)$ is the power spectrum at the i-th multipole bin of $\ell_i$,
$\mathrm{Cov}(\ell_i, \ell_j)$ is the covariance between $C(\ell_i)$ and $C(\ell_j)$,
and the sum in Eq.~(\ref{eq:s2n_power}) runs over all indices with $\ell_i \le \ell_\mathrm{max}$ and $\ell_j \le \ell_\mathrm{max}$.

We find that the convergence power spectrum predicted by our model is in agreement with the counterparts in the true lensing simulations in many aspects.
On the ensemble average, the prediction by our generators provides an accurate fit to the simulation counterpart as shown in the left panels in Figure~\ref{fig:comp_powerspec}.
Apart from the largest angular scales of $\ell\simlt 200$ (corresponding to $\simgt 1\, \mathrm{degree}$), 
our generators can reproduce the information about two-point correlation functions in the lensing simulations within a 1\%-level accuracy.
\ms{At $\ell\simlt 200$, we find 10\%-level differences of the power spectra between the ground truth and expected halofit predictions. These differences are left during the image-to-image translation by our GAN, because the GAN's outputs can hold the same large-scale modes as input GRFs. In fact, we observe that the cross correlation coefficients of the power spectra between the input Gaussian fields and GAN's outputs are consistent with unity at $\ell\simlt 200$. We speculate the
differences in the large-scale powers may be caused by a finite sky coverage of the true ray-tracing simulations.}

The variance in $C(\ell)$ by our generators is also consistent with the simulation counterpart in a wide range of multipoles.
We observe that our generators can enhance the variance at smaller angular scales as predicted by the ground truth.
Note that this enhancement arises from non-linear effects in gravitational growth of cosmic mass density \citep{Takada:2013wfa}, sourcing the non-Gaussanity in the lensing convergence.
Similar non-Gaussian contributions are also prominent in off-diagonal elements in the covariance matrix, reducing the cumulative S/N of $C(\ell)$ as small-scale information is added \citep{Sato:2009ct}. 
The right panels in Figure~\ref{fig:comp_powerspec} demonstrate that the power spectra produced by our generators preserve almost the same information contents as in 
the true simulation.
The degradation in the S/N with respect to the Gaussian expectation can be explained by our model as well.

It is beneficial to compare the covariance matrix of $C(\ell)$ by our model with the simulation counterpart in a direct manner because the covariance plays a central role in determining the constraining power of $C(\ell)$ for a given survey setup.
Figure~\ref{fig:comp_powerspec_cov} summarizes the covariance matrix predicted by our generators as well as the simulation counterparts.
Each panel represents the covariance $\mathrm{Cov}(\ell, \ell')$ as a function 
of $\ell$ at a fixed $\ell'$.
Note that we normalize the covariance so that its diagonal element can be unity.
For comparison, the predictions by the popular log-normal models are shown in
the orange dashed lines in the figure.
We find that the log-normal model can account for some levels of the non-Gaussian covariance, but its validity is still limited.
The log-normal model introduces a non-zero covariance at the smallest scales of $\ell\sim10^{4}$, but it cannot explain the cross-correlations between two $\ell$ bins when the difference in $\ell$ becomes large. 
Our neural-network-based model can account for significant covariance at a wide range of $\ell$ as seen in the true simulation data.
Note that the significant non-Gaussian covariance in the true lensing simulations 
can be mostly explained by statistical fluctuations 
in the number of dark matter halos sampled by a finite survey area, 
referred to as halo sample variance (HSV) \citep{Sato:2009ct, Takada:2013wfa}.
Our model can efficiently learn the HSV effects.

To get some sense of the learning experience of our model,
we examine to what extent our model performance can depend on the choice of the number of labels in the training datasets.
It is worth noting that we introduce the labeling of the training datasets by the field variance of lensing convergence fields as in Figure~\ref{fig:var_labels}.
This labeling ensures that the convergence field generated by our generators can have a similar field variance to the true counterpart.
Because the HSV effects in the covariance of $C(\ell)$ depend on the variance in cosmic mass density in a survey window, we expect that the field-variance-based labeling can make our model efficiently learn the HSV.
Figure~\ref{fig:comp_powerspec_varying_labels} shows the model performance when we vary the number of labels in the training datasets.
We observe that the training with no labels cannot appropriately teach the HSV effects.
The model trained without the field-variance-based labels cannot fully explain the non-Gaussian covariance at small angular scales which is a distinct feature in the true lensing simulations.
The labeling significantly improves in predicting the scatter in $C(\ell)$ at a wide range of $\ell$, indicating that the field variance is closely related to the HSV effects in the lensing convergence field as expected.
We also confirm that our model performance can be well converged as long as the number of labels is set to 10.
Our results highlight the importance of physically-motivated annotations in unsupervised training of neural networks for cosmological purposes.

\ms{The failure of predicting the covariance is interpretable that our GAN may suffer from the mode collapse. A more sophisticated framework \citep[e.g.][]{pmlr-v70-arjovsky17a} may help solving this issue, but we leave it for future research. It is also beneficial to develop a single generative model conditioned on the field-variance labels in order to reduce the training time. In our computation resource with NVIDIA A100 tensor core, we require about 100 hours to finish training genenerators with 10 different labels. A well-designed architecture accounting for label dependences may be able to reduce the training time by a factor of $\sim10$.}

\subsection{Bispectrum}

\begin{figure*}[!t]
\includegraphics[clip, width=2.1\columnwidth]{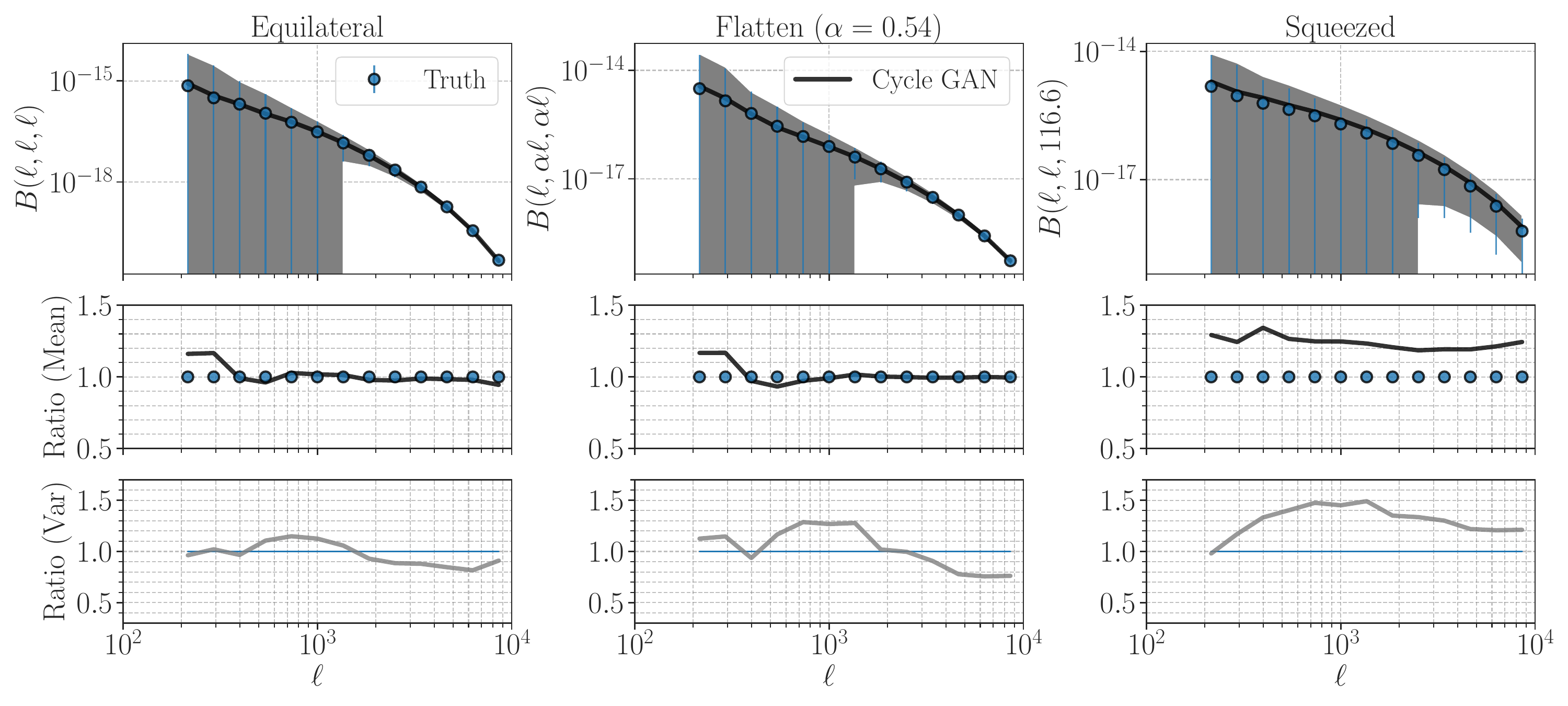}
\caption{\label{fig:comp_bispec} 
Comparison of convergence bispectra predicted by the lensing simulation and our model.
The left, middle, and right panels show the bispecra with three different configurations of equilateral, flattened, and squeezed shapes, respectively.
The blue circles with the error bars in the upper panels represent the simulation results, while the black solid lines stand for the model prediction.
The shaded region in the upper panels highlights the statistical uncertainty of the bispectra by our model.
In the second row, we show the ratio of the average bispectra between the two
(the black solid lines there show the model bispectra divided by the simulation counterparts). The bottom panels summarize the ratio of the variance.
The gray lines at the bottom are the model variance divided by the simulation counterparts. 
All the results are based on 10,000 realizations of the simulations and the neural network-based model. 
}
\end{figure*}

We study the lowest-order non-Gaussian statistic of the lensing bispectrum predicted by our neural network-based model.
The bispectrum provides information about three-point spatial correlations 
among several length scales. 
Because our model aims for an image-to-image translation and the input Gaussian convergence should have zero bispectrum on average, the bispectrum can be a good measure to look into the non-Gaussianity acquired through the translation by our model.

Figure~\ref{fig:comp_bispec} summarizes the average and variance of the bispectrum over 10,000 realizations of the fake convergence fields by our model.
In the figure, we consider three representative configurations of Fourier modes, 
i.e.~equilateral, flattened and squeezed shapes.
We find that our model can reproduce the equilateral and flattened shaped bispectra in the simulations with a level of $<10\%$ on average, while its variance is also consistent with the true simulation counterparts. 
It would be worth emphasizing that the popular log-normal model cannot provide a reasonable fit to the bispectrum in the true simulation as seen in Section~\ref{subsec:comp_lognormal}.
On the squeezed configuration, we observe that the model prediction deviates from the true counterparts with a level of $\sim30\%$ and $\sim50\%$ for the average and variance, respectively.
Note that the current state-of-the-art fitting formula of the matter bispectrum can explain the lensing bispectrum in simulations with a $10\%$-level deviation \citep{Takahashi:2019hth}.

Hence, our model learns the image-to-image translation so that it can preserve novel information about the three-point correlations in the lensing simulations.
This highlights that our model does not modulate convergence values from the input Gaussian one on a pixel-by-pixel basis.
Phase correlations in the Fourier space are a key ingredient for non-zero bispectrum \citep{Matsubara:2003te}, whereas there are no phase correlations in the input Gaussian model by construction (see Section~\ref{subsec:gaussian}).
The non-zero bispectra predicted by our model indicates that the image-to-image translation is operated in a multi-scale manner. 

\begin{figure*}[!t]
\includegraphics[clip, width=2.1\columnwidth]{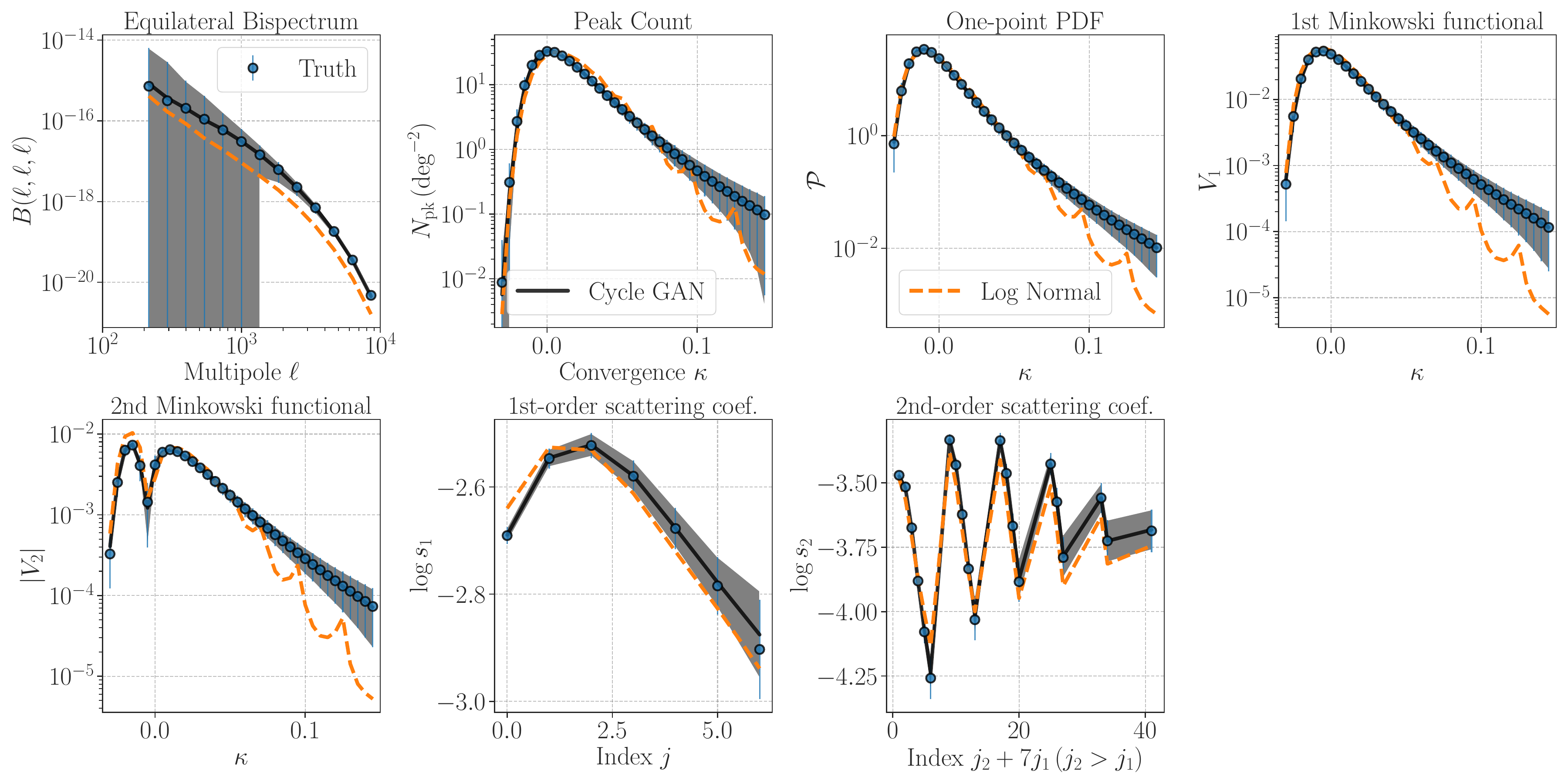}
\caption{\label{fig:comp_stats_lognormal} 
Convergence summary statistics predicted by different generative models.
Different seven panels summarize the statistics of the equilateral bispectrum, the peak count, the one-point PDF, the first MF ($V_1$), the second MF ($V_2$), and the ST coefficients up to second order ($s_1$ and $s_2$). 
In each panel, the blue circles with error bars show the average statistic and standard deviation obtained from the 10,000 lensing simulations. The black solid lines in the panels are our model predictions, and the gray-shaded regions stand for the standard deviation in our model.
For comparison, the orange dashed lines represent the average statistics estimated by 10,000 realizations of the log-normal model.
}
\end{figure*}

\begin{figure*}[!t]
\includegraphics[clip, width=2.1\columnwidth]{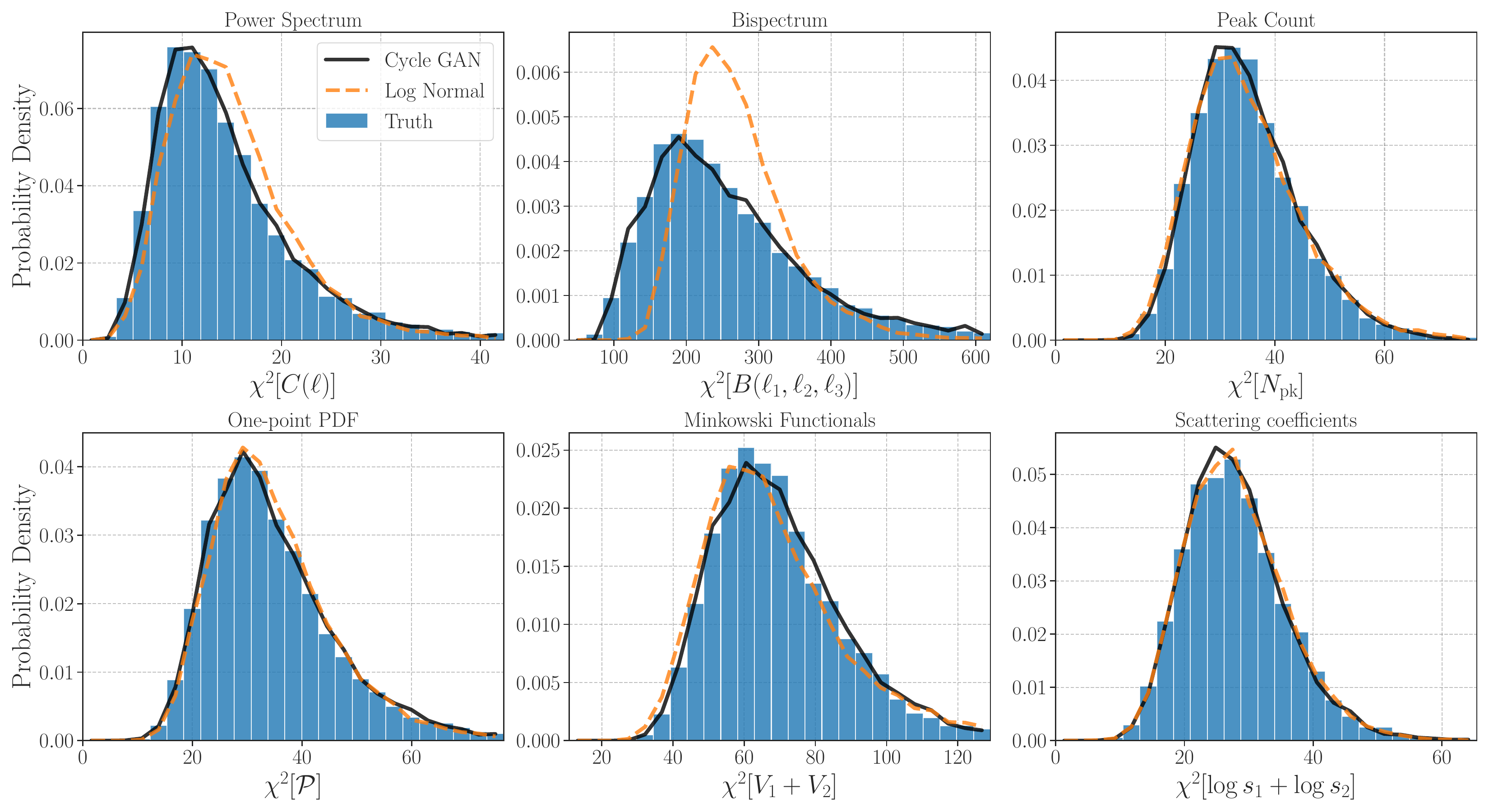}
\caption{\label{fig:comp_chisq} 
Distributions of chi-square quantities for various summary statistics.
The chi-square quantity for a given statistic $S$ is defined as in Eq.~(\ref{eq:chisq_realizations}). In each panel, the blue histogram shows the distribution obtained from 10,000 lensing simulations, while the black solid lines are our model predictions based on the neural networks.
For comparison, the orange dashed line in each panel represents the distribution for 10,000 log-normal realizations.
In this figure, we use all bins of statistics $S$ measured in a given convergence field with a sky coverage of $3.5^2\, \mathrm{deg}^2$. 
Details of binning are found in Section~\ref{subsec:summary_stats}.
It is worth noting that the results in this figure do not account for any observational effects.
}
\end{figure*}

\subsection{Comparison with log-normal model}\label{subsec:comp_lognormal}

In this section, we check the performance of our model to predict various summary statistics and compare the model predictions with the log-normal counterparts.
Figure~\ref{fig:comp_stats_lognormal} shows a set of summary statistics by our neural-network-based model, including the equilateral-shaped bispectrum, the peak counts,
the one-point PDF, the MFs, and the ST coefficients.
We see that the log-normal bispectrum is off from the simulation counterparts,
whereas our model can reproduce the equilateral bispectrum in the simulations in an accurate way. 
On the peak count, the one-point PDF, and the MFs, we find that our model improves 
in predicting statistical properties around rare convergence regions with a 
$\simgt5\sigma$ significance compared to the log-normal model. 
We also observe that the ST coefficients by our model are in good agreement with the simulation counterparts, implying that our model may be useful for cosmological inferences based on neural networks.

Importantly, we also find that our model can reproduce the variance of each statistic in the true simulations.
For instance, our model can predict the variance of the equilateral bispectrum 
over a wide range of $\ell$ with a 10\%-level accuracy, while
it can provide the variances of the peak counts, the one-point PDF, and the MFs in the range of $0 \simlt \kappa \simlt 0.15$ with a 20-30\% level accuracy.
We also observe that the labeling of the training datasets with the field variance is effective in predicting the variance of various summary statistics. 
When we trained our model without introducing any labels, the model tended to underestimate the variance of the peak counts, the one-point PDF, and the MFs at a wide range of $\kappa$ by $50\%$ or larger.

To compare our model predictions 
with the log-normal counterparts in a quantitative way, 
we introduce a chi-square quantity defined as
\beqa
\chi_{r}^2[S] = \sum_{i,j} 
(S^{(r)}_i - \bar{S}_i) \, \mathrm{Cov}^{-1}_{ij}[S]\, (S^{(r)}_j-\bar{S}_{i}), \label{eq:chisq_realizations}
\eeqa
where $\bar{S}_i$ represents the ensemble average at the i-th bin of a given binned summary statistic $S$, 
$\mathrm{Cov}_{ij}[S]$ is the covariance matrix for the binned $S$,
and $S^{(r)}_{i}$ is the i-th summary statistic at the $r$-th realization.
We evaluate the ensemble averages and covariance matrices for various summary statistics over 10,000 realizations of the true lensing simulations as well as the fake convergence fields generated by our model and the log-normal counterparts.
We then compute 10,000 chi-square quantities for the three generative models 
and compare the histogram of $\chi^2[S]$ among the different models.
The comparisons are important in practice because the chi-square quantity in Eq.~(\ref{eq:chisq_realizations}) commonly provides an estimate of log-likelihood for the statistic $S$, which is a central part of modern cosmological analyses.

Figure~\ref{fig:comp_chisq} compares the histograms of $\chi^2[S]$ for six different sets of summary statistics $S$ among the three cases of the true lensing simulations, the log-normal model, and our neural-network-based model.
We find that our model can reproduce the distribution of $\chi^2$ in the true simulations for every statistic, while the log-normal counterpart exhibits notable differences from the true one for some statistics.
The clearest difference between the true simulations and the log-normal model is found in the distribution of $\chi^2$ for the bispectrum.
Apart from the bispectrum, we observe that the distribution by our model explains the median in the true $\chi^2$ distribution better than the log-normal counterpart.
Note that we do not include any observational effects in the comparison, indicating that the results in Figure~\ref{fig:comp_chisq} do not provide a realistic log-likelihood function for given statistics.
Also, further study is needed to validate if GAN-based log-likelihood computations can be used for parameter inference in a realistic setup.



\section{Extension of Sky Coverage}\label{sec:extension_sky}

For an application of our generative model, we here examine to produce a weak lensing mass map covering a continuous sky larger than the field-of-view of individual training data. Note that previous neural-network-based models in the literature \citep{2019ComAC...6....1M, Perraudin:2020gig, Remy:2022ixn} cannot produce a continuous map if the sky coverage becomes larger than the one assumed in the training datasets. Considering that our model utilizes the Gaussian model as latent information, we propose a four-step procedure to generate a non-Gaussian convergence field with an arbitrary sky coverage;

\begin{enumerate}
    \item[(i)] Create a Gaussian realization of the convergence field with a given sky coverage $\Omega_\mathrm{large}$,
    \item[(ii)] Chop the Gaussian convergence field into a set of subregions so that each subregion can cover the field-of-view of single training datasets, denoted as $\Omega_\mathrm{small} = 3.5^2\, \mathrm{deg}^2$,
    \item[(iii)] Conduct an image-to-image translation at each subregion with our generators,
    \item[(iv)] Concatenate the non-Gaussian fields generated in the previous step (iii).
\end{enumerate}

We then summarize some additional treatments on a step-by-step basis.

\begin{figure*}[!t]
\includegraphics[clip, width=2.1\columnwidth, rviewport= 0.05 0.05 1.0 0.9]
{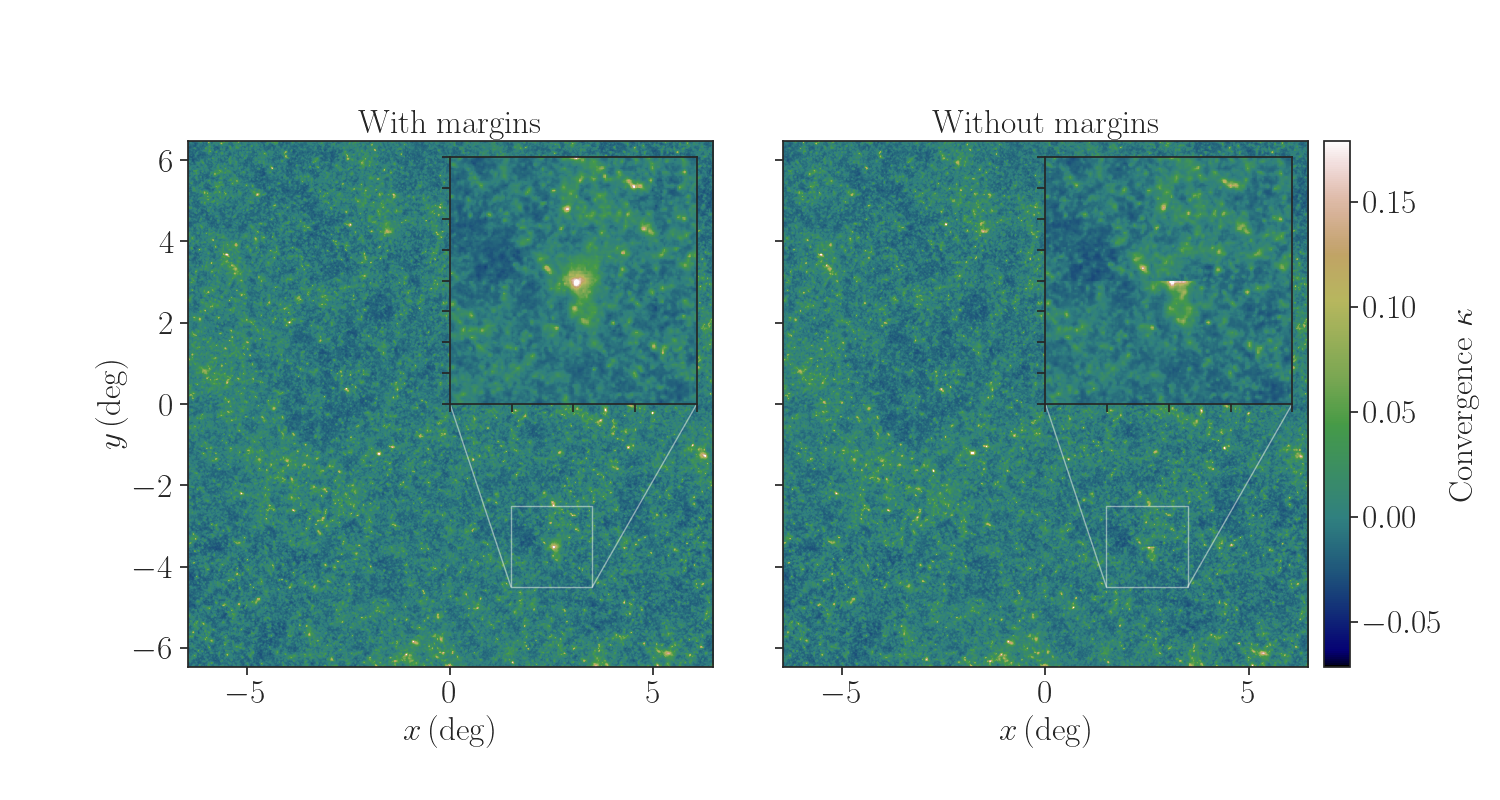}
\caption{\label{fig:cycle_GAN_largesky_example} 
An example of extension of sky coverage with our neural-network-based generators. The left panel shows our baseline method including margins around boundaries of small subregions, while the result without margins is presented on the right. In each panel, the inset figure highlights a zoom-in view around the high convergence value at the boundary between two subregions.
}
\end{figure*}


\paragraph*{Step (i)}
Because our neural-network-based model needs an input image with the size of $256\times256$, we set the size of a large-sky Gaussian convergence field to 
$(256 M_x)\times(256 M_y)$ where $M_x$ and $M_y$ are integers. We consider $M_x=M_y=4$ for the demonstration below. Note that the pixel size of $0.82\, \mathrm{arcmin}$ on a side is required to be the same as in the training datasets. Hence, a single Gaussian convergence field produced in this step covers a sky of $\Omega_\mathrm{large}=14^2 \, \mathrm{deg}^2$.

\paragraph*{Step (ii)} When dividing the large-sky Gaussian convergence field into small subregions, we decompose it into two terms of
\beqa
\kappa_\mathrm{G, large}(\bm{\theta}) = 
\kappa^{(L)}_\mathrm{G, large}(\bm{\theta}) + 
\kappa^{(S)}_\mathrm{G, large}(\bm{\theta}),
\eeqa
where $\kappa_\mathrm{G, large}$ represents the Gaussian convergence field produced at step (i), and the two fields on the right-hand side are produced by high- and low-pass filtering of $\kappa_\mathrm{G, large}$ such that
\beqa
\kappa^{(L)}_\mathrm{G, large}(\bm{\theta}) &\equiv& \int\, \mathrm{d}^2\theta' W^{(L)}(\bm{\theta}-\bm{\theta}')\, \kappa_\mathrm{G, large}(\bm{\theta}'), \\
\kappa^{(S)}_\mathrm{G, large}(\bm{\theta}) &\equiv& \int\, \mathrm{d}^2\theta' W^{(S)}(\bm{\theta}-\bm{\theta}')\, \kappa_\mathrm{G, large}(\bm{\theta}').
\eeqa
In the above, we introduce the high- and low-pass filters of $W^{(L)}$ and $W^{(S)}$, respectively.
In this paper, we define those two filters to be top-hat filters in Fourier space.
To be specific, each filter in Fourier space is given by
\beqa
\tilde{W}^{L}(\bm{\ell}) &=& {\cal H}(\ell_\mathrm{cut}-|\bm{\ell}|), \\
\tilde{W}^{S}(\bm{\ell}) &=& 1-\tilde{W}^{L}(\bm{\ell}),
\eeqa
where ${\cal H}(x)$ is the Heaviside step function which is 1 for $x>0$ and 0 otherwise,
and $\ell_\mathrm{cut}$ is a cutoff multipole.
We calibrate $\ell_\mathrm{cut}$ so that the field variance of 
$\langle [\kappa^{(S)}_\mathrm{G, large}]^2 \rangle$ over a $3.5\, \mathrm{deg}^{2}$ sky can follow the same distribution as in the Gaussian simulations used for the training.
This calibration is important to avoid covariate shifts \citep{SHIMODAIRA2000227}, ensuring that input Gaussian images for our generators should have the same statistical nature as the training data. We find that $\ell_\mathrm{cut} = 0.55 \ell_\mathrm{f}$ provides a good fit, where $\ell_\mathrm{f} = 2\pi/\Omega^{1/2}_\mathrm{small} = 102.8$ represents the fundamental Fourier mode in the training datasets.

We then divide the field of $\kappa^{(S)}_\mathrm{G, large}$ into $M_x \times M_y$ subregion so that each subregion can cover the sky coverage of $\Omega_\mathrm{small}$.
We set the number of pixels to be $256\times256$ in each subregion, but 
we extract each subregion with a 20-pixel overlap region around the edges of every subfield, ensuring that the final map is continuous after combining subfields \citep{Han:2021unz}.
Note that the overlap regions are not used to produce the final map.
Hence, this procedure removes $(M_x\times20)\times(M_y\times20)$ pixels in our map making, leading the final map to cover a sky coverage of $12.9^2\, \mathrm{deg}^2$.
We hereafter define $\kappa^{(S)}_\mathrm{G, small}$ as the high-pass filtered Gaussian field with a sky coverage of $\Omega_\mathrm{small}$.

\begin{figure*}[!t]
\includegraphics[clip, width=2.1\columnwidth]{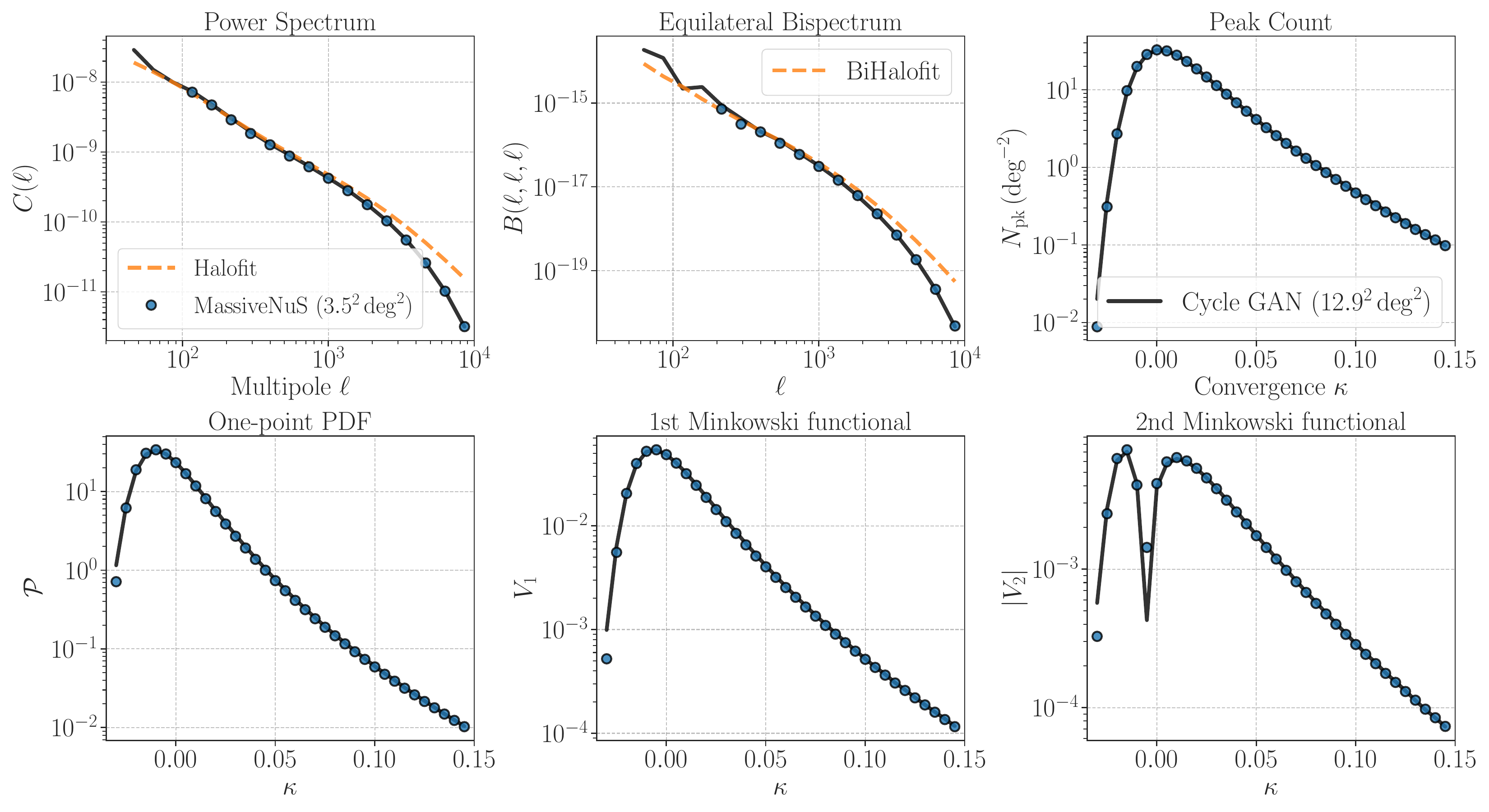}
\caption{\label{fig:comp_avg_largesky} 
Comparison of summary statistics for lensing convergence fields 
with different sky coverages. Six panels show the ensemble average of different summary statistics including power spectrum, equilateral-shaped bispectrum, peak counts, one-point PDF, and two Minkowski functionals.
In each panel, the blue points represent the average summary statistics 
over 10,000 realizations of the true lensing simulations, whereas the black solid line stands for the average obtained from 500 realizations of the fake convergence fields generated by our neural network.
Each lensing simulation covers a $3.5^2\, \mathrm{deg}^2$, and the fake field assumes a larger sky coverage of $12.9^2\, \mathrm{deg}^2$.
For the power spectrum and the bispectrum, we overlay theoretical predictions based on Eqs.~(\ref{eq:kappa_power}) and (\ref{eq:kappa_bispec}) as shown in the orange dashed lines. To compute the theoretical predictions, we use the fitting formula of non-linear matter power spectrum and bispectrum developed in \citet{Takahashi:2012em, Takahashi:2019hth}.
}
\end{figure*}

\paragraph*{Step (iii)}
Given the $M_x \times M_y$ fields of $\kappa^{(S)}_\mathrm{G,small}$,
we add non-Gaussian features to individual fields by our generators.
This process is formally written as
\beqa
\kappa^{(S)}_\mathrm{NG, small}(\bm{\theta}) = f_\mathrm{NN}\left[\kappa^{(S)}_\mathrm{G,small}(\bm{\theta}) | \sigma^2_\mathrm{G,small}\right],
\eeqa
where $f_\mathrm{NN}(\cdot|\sigma^2_\mathrm{G, small})$ represents our neural-network-based model conditioned by the field variance of $\sigma^2_\mathrm{G, small}$.
For the field variance, we compute $\sigma^2_\mathrm{G, small}$ as 
$\langle [\kappa^{(S)}_\mathrm{G,small}]^2\rangle$ over each subregion.
It would be worth noting that we trained 10 generators according to the field variance of the input Gaussian fields. Hence, we have 10 different models of $f_\mathrm{NN}$ in this paper.

\paragraph*{Step (iv)}
We then combine the fields of $\kappa^{(S)}_\mathrm{NG, small}$ after removing the 20-pixel overlap regions so that a single combined field covers a $12.9^2\, \mathrm{deg}^2$ sky.
In the joining process, we also add the low-pass filtered Gaussian field $\kappa^{(L)}_\mathrm{G, large}(\bm{\theta})$ in order to preserve the large-scale information inherent with the latent field of $\kappa_\mathrm{G, large}(\bm{\theta})$.

\ms{We first examine the effect of margins 
when combining small-area fields into a single large-area field in Figure~\ref{fig:cycle_GAN_largesky_example}.
The left panel shows the extenstion of sky coverage by our baseline method with the 20-pixel margins, while the right represents a naive extenstion without any margins. The figure clearly demonstrates that the overlapping regions around each subregion are important to remove an undesired discontinuity.}

Figure~\ref{fig:comp_avg_largesky} shows ensemble averages of various lensing statistics for the non-Gaussian convergence field covering a sky of $12.9^2\, \mathrm{deg}^2$.
In the figure, we summarize the results obtained from 500 realizations of the large-sky non-Gaussian convergence fields by following the steps (i)-(iv).
Note that it takes about 60 seconds to produce a large-sky map with $M_x=M_y=4$ with the use of a single GPU\footnote{We use NVIDIA A100 tensor core with a memory size of 40GB throughout this paper.}.
The figure highlights that our model enables us to reproduce the lensing summary statistics as same as the true lensing simulations even if extending a sky coverage.
The upper left panel shows the average power spectra with two different sky coverages.
For comparison, the orange dashed line in the upper left represents a prediction by Eq.~(\ref{eq:kappa_power}).
On the large scale at $\ell \simlt 100$, our model does not learn any information about the power spectrum in the training process, but it can provide a good fit to the theoretical prediction.
A similar result can be found in the upper middle panel showing the equilateral bispectrum.
We also observe that our model is still able to reproduce the ensemble averages of peak counts, one-point PDF, and the MFs as in the true lensing simulations when the sky coverage of the model is set to be larger than the counterpart of the training data. 
Note that the peak counts and MFs can be affected by possible discontinuity across the subregions in our map-making because those statistics are defined with the information of spatial derivatives in the lensing convergence. 
The results in Figure~\ref{fig:comp_avg_largesky} demonstrate that extension of the sky coverage based on steps (i)-(iv) works with no practical problems.

\section{Limitation}\label{sec:limitation}

Before concluding this paper, we summarize the major limitations of 
our neural network-based model. The following issues will be addressed in future studies.

\subsection{Squeezed bispectrum}
Although our model can reproduce most of the lensing summary statistics as in the ground truth, there is room for further improvements in the model performance.
We find that the squeezed-shaped bispectrum is one of the most difficult 
statistic to accurately predict our current model. 
The squeezed bispectrum is commonly defined by the bispectrum 
$B(\bm{\ell}_1, \bm{\ell}_2, \bm{\ell}_3)$ with one multipole being significatly smaller than other multipoles.
In the end, the number of triangles at the squeezed configurations 
is limited in our training datasets.
To improve the model prediction of the squeezed bispectrum without increasing the size of training datasets, one may consider a weakly non-Gaussian convergence field as the input of our generators instead of the simplest Gaussian model.

For a given theoretical template of the bispectrum $B_\mathrm{ref}(\bm{\ell}_1, \bm{\ell}_2, \bm{\ell}_3)$, one can construct the non-Gaussian field so that its average bispectrum can be equal to the theoretical counterpart by \citep{Smith:2006ud,Hall:2017pzf}
\beqa
\tilde{\kappa}_\mathrm{NG, pert}(\bm{\ell})
&=& \tilde{\kappa}_\mathrm{G}(\bm{\ell})
+ \frac{1}{6}\int \frac{\mathrm{d}^2\ell'}{(2\pi)^2}\, B_\mathrm{ref}(\bm{\ell}, -\bm{\ell}', -\bm{\ell}+\bm{\ell}') \nonumber \\
&&
\quad \quad \quad \quad \quad \quad 
\times
\frac{\tilde{\kappa}_\mathrm{G}(\bm{\ell}') \tilde{\kappa}_\mathrm{G}(\bm{\ell}-\bm{\ell}')}{C_\mathrm{G}(\ell')C_\mathrm{G}(|\bm{\ell}-\bm{\ell}'|)}, \label{eq:non-G-kappa}
\eeqa
where $\tilde{\kappa}_\mathrm{G}(\bm{\ell})$ represents a Gaussian convergence in Fourier space with zero mean and its average power spectrum being $C_\mathrm{G}$.
The bispectrum of $\kappa_\mathrm{NG, pert}$ is equal to the reference 
$B_\mathrm{ref}$ on average by construction. 
Once the squeezed bispectrum template can be obtained by numerical simulations and/or analytic approaches, one can include the information with Eq.~(\ref{eq:non-G-kappa}) in the training datasets.
This pre-training would be among the simplest approaches to update our neural network-based model in a physically intuitive way, but detailed examinations are required.  

\subsection{Curved Sky}
Throughout the paper, we assume a flat sky to generate weak lensing mass maps, 
whereas the assumption can break as the survey area increases shortly \citep{Wallis:2017lwt, Kansal:2023jkg}. 
The flat-sky approximation is valid for the standard power-spectrum analysis in most cases \citep{Kitching:2016zkn, Kilbinger:2017lvu, Lemos:2017arq}, 
but the validity would depend on sky positions for other non-Gaussian statistics.
For instance, projections of a sphere on a plane can induce some distortions of the surface, affecting precise estimates of peaks and contours in the lensing convergence.
A promising approach to account for curved-sky effects is to upgrade the architecture of our neural networks with a graph convolutional neural network as introduced by \citet{Perraudin:2018rbt}.
However, it seems difficult to learn a sphere-to-sphere translation by keeping the angular resolution to sub-arcminutes, because the number of grids on an input sphere 
can be of an order of $O(10^{8})$ in the common HEALPix pixelization \citep{Gorski:2004by}. 
A well-organized training strategy on a patch-by-patch basis is needed in practice \citep{Han:2021unz}. 

\subsection{Generative models for lensing tomography}
Modern weak lensing analyses commonly employ a tomographic setup 
consisting of multiple lensing convergence fields 
with different source redshift information.
Although it is easy to increase the number of channels of inputs and outputs 
for our neural networks, we expect that appropriate labels for multiple-channel training datasets would be important to make our generative model applicable to realistic tomographic setups.
There are several candidates for the label 
even when we assume input images to be produced by a multivariate Gaussian model.
The candidates include the field covariance matrix $\langle \kappa_\mathrm{G, i} \kappa_\mathrm{G,j} \rangle$ where $\kappa_\mathrm{G, i}$ represents the Gaussian convergence field at the i-th tomographic bin and the average is computed as the spatial average over a field-of-view in individual training datasets.
We assume that the field covariance can be a good measure to quantify the HSV effects in the tomographic lensing analysis as motivated by our single-channel analysis, 
but this assumption should be examined with actual simulation data.
Note that the labeling of the field covariance cannot be determined uniquely, indicating that some trial and error would be crucial.

\subsection{Dependence of cosmological and astrophysical parameters}
The ultimate goal in the generative model of weak lensing mass maps is 
to rapidly-produce a map for various cosmological models 
while accounting for different astrophysical effects such as source redshift distributions, intrinsic alignments \citep{Troxel:2014dba}, and baryonic effects in the cosmic mass density \citep{Chisari:2019tus}.
Our study can provide an important first step toward such an all-in-one generative model, but it is still challenging to add realistic cosmological and astrophysical dependence to a deep-learning-assisted simulator in general.
There are some attempts to develop a cosmology-dependent generative model of weak lensing or two-dimensional mass density maps in the literature \citep{Perraudin:2020gig, Tamosiunas:2020rvw, Dai:2022dso}.
It would be worth noting that previous cosmology-dependent generators limit themselves to work with a fixed sky coverage, questioning their utilities in a real-world analysis.
Few-shot image generation \citep[e.g.][]{9577580} is among the most promising approaches to update our model so that it can be adapted to 
various cosmological and astrophysical scenarios.
For instance, a baseline generative model can be trained with a large set of fiducial training data, while few-shot learning allows one to update the baseline generator with a small set of training data at different cosmological and astrophysical models.
The few-shot learning already has been examined in a classification problem of galaxies \citep{2022RAA....22e5002Z}, and it might be effective for image generation tasks in astronomy.

\section{Conclusion}\label{sec:conclusion}

In this paper, we have proposed 
a new generative model of the weak lensing mass maps (convergence fields) based on an image-to-image translation by neural networks. 
The model aims at translating Gaussian convergence fields to fully non-Gaussian counterparts as simulated by cosmological ray-tracing simulations \citep{Liu:2017now}.
We have trained the neural networks within 
the framework of Cycle-Consistent Generative Adversarial Networks (Cycle GAN) \citep{CycleGAN2017},
enabling the networks to learn the style transfer between two different convergence models in an unsupervised manner.
We have found that annotating the training datasets is key 
to realizing the desired level of diversity in the convergence fields 
generated by our model.
Labeling individual training images with field variances is 
effective in learning non-Gaussian covariance structures in the convergence power spectrum for our neural network-based model.

Compared with a popular log-normal convergence model, 
our model can provide more accurate predictions of convergence bispectra as well as local structures around rare high-density regions.
We have validated our model by studying multiple sets of convergence summary statistics, including the bispectra, number density of convergence peaks, 
one-point probability distribution function, Minkowski functionals, and scattering transform coefficients.
The model predictions of those summary statistics are statistically consistent with the counterparts in the true lensing simulations.
Likelihood functions of the model summary statistics 
can give a reasonable fit to the ground truth, highlighting that our generative model can play a central role in validating pipelines for cosmological inference.

We have also examined to produce a non-Gaussian convergence field with an arbitrary sky coverage by our neural network-based model.
For this purpose, we have laid out a four-step procedure; 
(i) Generate a Gaussian convergence map, (ii) Divide the Gaussian map into small patches, (iii) perform an image-to-image translation on a patch-by-patch basis, and (iv) combine the resultant non-Gaussian fields in all the patches.
We have demonstrated that the four-step procedure works in producing a non-Gaussian convergence field with a sky coverage of $\sim166\, \mathrm{deg}^2$ in a GPU-minute, whereas our neural networks have been trained with the simulations covering a sky of $\sim12\, \mathrm{deg}^2$.
We have confirmed that the model summary statistics are still consistent with the ground truth even when extending sky coverage.
This method opens a valuable opportunity for massive productions of weak lensing mass maps with realistic non-Gaussianities but covering large areas of the sky.
Note that our trained generators only require low-cost Gaussian convergence fields to produce a new set of realistic non-Gaussian counterparts, allowing us to increase the number of independent non-Gaussian convergence fields in a rapid manner.

Our generative model accelerates the production of mock weak lensing mass maps in a simple and efficient way, but it has to be revised in various aspects for real-world applications.
The model tends to overestimate a squeezed-shaped bispectrum 
with a level of $30\%$, indicating that the input convergence field for our model should preserve weakly non-Gaussian structures.
We constructed our generative model assuming single source redshift in a flat sky, but this is not valid for future lensing surveys with a sky coverage larger than $1000\, \mathrm{deg}^2$ and tomographic setups.
Our model is not able to produce a weak lensing mass map as a function of various cosmological and astrophysical parameters, which is a common weak point in deep-learning-assisted approaches.
We expect that the few-shot learning method can be an effective breakthrough in cosmology- and astrophysics-dependent generators, 
as a set of various simulation data becomes available.
This is along the lines of our future study.

\begin{acknowledgments}
We thank Naoki Yoshida and Kana Moriwaki for the useful comments.
This work is supported by MEXT KAKENHI Grant Numbers 19K14767 and 20H05861.
A part of numerical computations was carried out on Cray XC50 at the Center for Computational Astrophysics in NAOJ.
\end{acknowledgments}


\bibliography{refs}

\appendix
\section{Network Architecture in Cycle GAN}\label{apdx:network}

In this appendix, we provide some details of the architecture for our generative networks. In our Cycle GAN, we adopt the same architecture for a pair of GANs.
For the generator, we follow the architecture designed by \citet{10.1007/978-3-319-46475-6_43}.
This network contains three convolutions, nine residual blocks \citep{7780459}, 
two transposed convolutions, and one convolution that generates $256\times256$ images.
As in \citet{10.1007/978-3-319-46475-6_43}, we insert an instance normalization module \citep{8099920}, improving the quality of image stylization.
For the discriminator, we adopt $70\times70$ PatchGANs \citep{8100115, 8099502, 10.1007/978-3-319-46487-9_43}, aiming at authenticity determination across
70 × 70 overlapping image patches. 
Note that a patch-level discriminator architecture allows us to reduce the number of parameters compared to a full-image discriminator \citep{8100115}.
Also, it can be implemented on arbitrarily sized images in a fully convolutional fashion,
making the training efficient with GPU environments.
Details of the architecture are provided in Tables~\ref{tab:Generator} and \ref{tab:Discriminator}.

\begin{table}[!t]
\caption{\label{tab:Generator}%
Generator network architecture: layer types, activations, 
output shapes (channels $\times$ height $\times$ width) and the number of trainable parameters for each layer. 
Conv($k$, $s$) are $k \times k$ Convolution layers with strides=$s$, while TConv($k$, $s$) stands for the transposed counterpart. 
ResNetBlock denotes a residual block that contains two 
$3\times3$ convolutional layers with the same number of filters on both layers.
ReflectionPad represents a reflection padding used for the reduction of artifacts.
InstanceNorm is the instance normalization.
}
\begin{ruledtabular}
\begin{tabular}{llll}
\textrm{}&
\textrm{Activ.}&
\textrm{Output shape}&
\textrm{Params.}\\
\colrule
Input & -- & [1, 256, 256] & -- \\
ReflectionPad & -- & [1, 262, 262] & -- \\
Conv(7,1) & -- & [64, 256, 256] & 3,200 \\
InstanceNorm & ReLU & [64, 256, 256] & -- \\
Conv(3,2) & -- & [128, 128, 128] & 73,856\\
InstanceNorm & ReLU & [128, 128, 128] & -- \\
Conv(3,2) & -- & [256, 64, 64] & 295,168\\
InstanceNorm & ReLU & [256, 64, 64] & -- \\
ResNetBlock & -- & [256, 64, 64] & 1,180,160 \\
ResNetBlock & -- & [256, 64, 64] & 1,180,160 \\
ResNetBlock & -- & [256, 64, 64] & 1,180,160 \\
ResNetBlock & -- & [256, 64, 64] & 1,180,160 \\
ResNetBlock & -- & [256, 64, 64] & 1,180,160 \\
ResNetBlock & -- & [256, 64, 64] & 1,180,160 \\
ResNetBlock & -- & [256, 64, 64] & 1,180,160 \\
ResNetBlock & -- & [256, 64, 64] & 1,180,160 \\
TConv(3,2) & -- & [128, 128, 128] & 295,040\\
InstanceNorm & ReLU & [128, 128, 128] & -- \\
TConv(3,2) & -- & [64, 256, 256] & 73,792 \\
InstanceNorm & ReLU & [64, 256, 256] & -- \\
ReflectionPad & -- & [1, 262, 262] & -- \\
Conv(7,1) & Tanh & [1, 256, 256] & 3,137 \\
\colrule
& & Total params & 11,365,633 
\end{tabular}
\end{ruledtabular}
\end{table}

\begin{table}[!b]
\caption{\label{tab:Discriminator}%
Similar to Table~\ref{tab:Generator}, this table shows the discriminator network architecture. In the discriminator, we set every LeakyReLU’s leakine to be 0.2.
}
\begin{ruledtabular}
\begin{tabular}{llll}
\textrm{}&
\textrm{Activ.}&
\textrm{Output shape}&
\textrm{Params.}\\
\colrule
Input & -- & [1, 256, 256] & -- \\
Conv(4,2) & LReLU & [64, 128, 128] & 1,088 \\
Conv(4,2) & -- & [128, 64, 64] & 131,200\\
InstanceNorm & LReLU & [128, 64, 64] & -- \\
Conv(4,2) & -- & [256, 32, 32] & 524,544\\
InstanceNorm & LReLU & [256, 32, 32] & -- \\
Conv(4,1) & -- & [512, 31, 31] & 2,097,664\\
InstanceNorm & LReLU & [512, 31, 31] & -- \\
Conv(4,1) & -- & [1, 30, 30] & 8,193 \\
\colrule
& & Total params & 2,762,689 
\end{tabular}
\end{ruledtabular}
\end{table}

\section{Translation from non-Gaussian convergence to the Gaussian counterpart}\label{apdx:RT2Gauss}

\begin{figure*}[!t]
\includegraphics[clip, width=2.1\columnwidth]{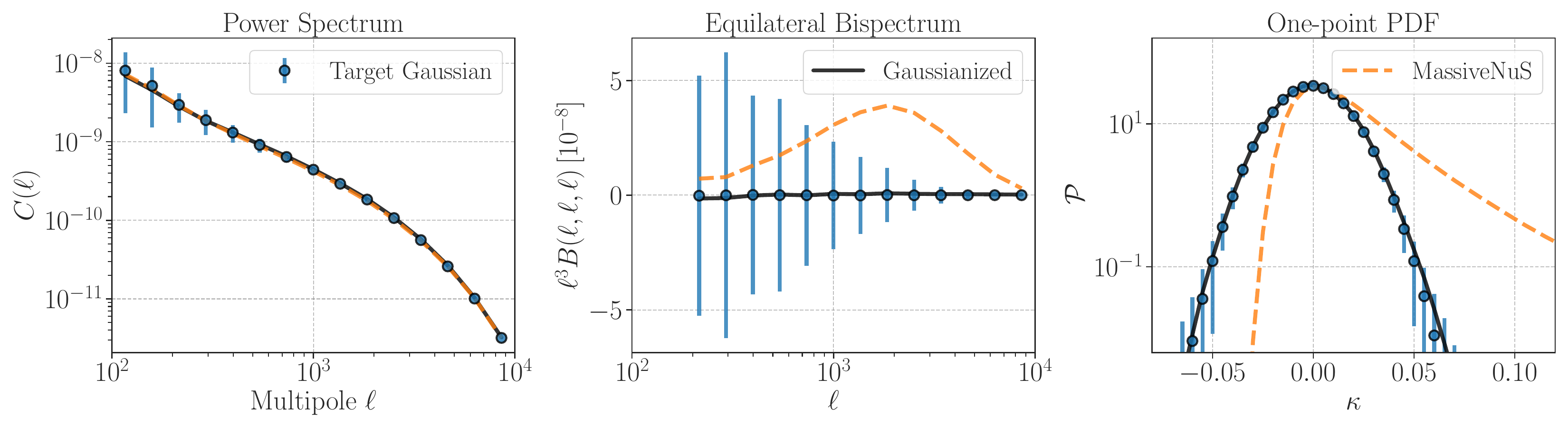}
\caption{\label{fig:comp_avg_gaussianized}
Comparisons of lensing summary statistics predicted by our neural-network-based ``Gaussianizer" and target Gaussian counterparts.
The left, middle, and right panels summarize the comparisons of power spectra, equilateral-shaped bispectra, and one-point PDFs, respectively.
In each panel, the blue points show the average statistics over 10,000 realizations of the target Gaussian generative model,
whereas the black solid lines stand for the prediction by our Gaussianizer.
The blue error bars in each panel correspond to the statistical uncertainty for a survey of $3.5\, \mathrm{deg}^2$.
For ease of comparison, we also include the average summary statistics over 
10,000 non-Gaussian convergence fields in the MassiveNuS simulation \citep{Liu:2017now}
as shown by the orange dashed lines in each panel.
This figure clarifies that our neural network can be used for a precise Gaussianization
of non-Gaussian convergence fields.
}
\end{figure*}

We here summarize the performance of the generator translating from non-Gaussian convergence maps into the Gaussian counterparts (i.e. Gaussianization), 
that is an interesting by-product of our Cycle GAN.
In this appendix, we show the results of the neural network trained without any labeling.

\begin{figure}[!t]
\includegraphics[clip, width=0.8\columnwidth]{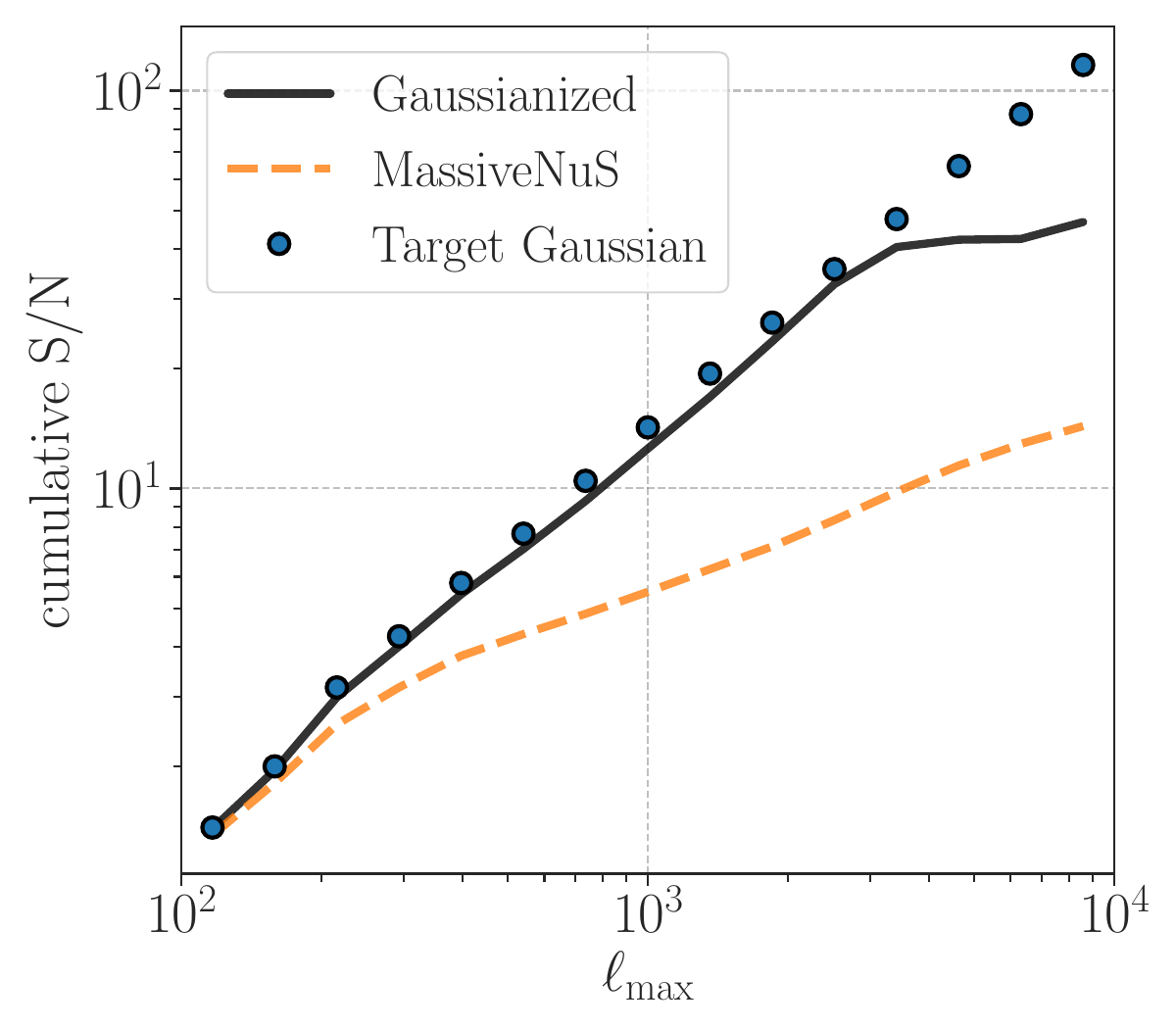}
\caption{\label{fig:comp_powerspec_s2n_gaussianized} 
Cumulative signal-to-noise ratios (S/Ns) of lensing power spectra for three different lensing maps. The cumulative S/N is defined in Eq.~(\ref{eq:s2n_power}).
The blue points represent the results of the target Gaussian maps, while the black solid and orange dashed lines stand for the Gaussianized and input non-Gaussian maps, respectively. A perfect Gaussianization ensures that the blue points and the solid line should be consistent with each other.
}
\end{figure}

Figure~\ref{fig:comp_avg_gaussianized} compares three different summary statistics between the target Gaussian and our generative model.
We chose power spectra, equilateral-shaped bispectra, and one-point PDFs for the comparison because these are primarily important to check 
the degree of Gaussianity left in a given lensing map.
The figure shows that our neural-network-based Gaussianization works well 
as long as the network was sufficiently trained.
It is worth noting that our Gaussianization vanishes the non-zero three-point correlations found in an input non-Gaussian lensing map.
This is an improvement compared to the previous Gaussianization methods \citep{YuYu:2011ijg,2011ApJ...729L..11S,Seo:2011ku,Simpson:2015nva},
because the previous approaches rely on a local transformation of non-Gaussian weak lensing mass maps in real space, making it difficult to erase the three-point correlations.

In Figure~\ref{fig:comp_powerspec_s2n_gaussianized},
we study the degree of Gaussianity in power spectra produced by our generator in detail.
We compute the cumulative signal-to-noise ratio (S/N) as in Eq.~(\ref{eq:s2n_power})
for the target Gaussian, the input non-Gaussian, and the Gaussianized maps.
If our generator can perform a perfect Gaussianization, the cumulative S/N should match the target Gaussian counterpart.
The figure highlights that our neural network-based Gaussianization cannot be perfect.
We observe that the cumulative S/N of the Gaussianized power spectrum deviates from the true Gaussian counterpart at $\ell\simgt3000$.
This means that the Gaussianization by our generator remains at some level of non-Gaussianity at angular scales of $\simlt 0.1\, \mathrm{deg}$.

We also caution that our Gaussianization assumes a fiducial cosmology for the image-to-image translation.
Hence, it is still uncertain if the Gaussianization can work even for non-Gaussian lensing maps at various cosmological models.
Another caveat in our method is that our image-to-image translation does not account for the presence of observational noises, which is the main practical difficulty in the Gaussianization of weak lensing fields.

\end{document}